\documentclass[
 reprint,
 amsmath,amssymb,
 aps,
 prb
]{revtex4-2}

\usepackage{bibunits}
\defaultbibliographystyle{apsrev4-2}
\defaultbibliography{references_andreev_molecules}
\usepackage{graphicx}
\usepackage{siunitx}
\usepackage[colorlinks=true, urlcolor=blue, linkcolor=blue, citecolor=blue, hyperfootnotes=false]{hyperref}

\newcommand{\bra}[1]{\langle #1 \rvert}
\newcommand{\ket}[1]{\lvert #1 \rangle}

\graphicspath{{Figures/}}

\begin{document}

\widetext

\title{Andreev molecules at distance}
\author{Erik S.~Samuelsen}
\email{e.s.s.samuelsen@tudelft.nl}
\author{Yuli V.~Nazarov}
\affiliation{Kavli Institute of Nanoscience, Delft University of Technology, 2628 CJ Delft, The Netherlands}

\begin{abstract}

Andreev molecule states arise from hybridization of Andreev bound states in different Josephson Junctions. Extensive theoretical and experimental research concentrates on direct coherent electron coupling between the junctions: this implies the distance between the junctions is of the order of superconducting coherence length, that is, short.   

We propose and discuss the possibility to create Andreev molecules at long (in principle, arbitrary long) distance  between the junctions. In this case, the hybridized states are excited quasi-particle singlets and the coupling is provided by an embedding electric circuit. To achieve a strong hybridization, one aligns the energies of the Andreev bound states with associated phase differences.  In fact, a recent experiment realizes such setup. 

With circuit theory we derive the hybridization level splitting and estimate the scale of the effect. Since the phenomenon encompasses excited states, we derive and solve the associated Lindblad equation under condition of persistent resonant excitation. By analyzing the resulting dissipative dynamics we identify relevant regimes where the hybridization and resonant excitation peaks are most pronounced. The low-frequency mutual inductance of the Josephson junctions is an important signature of the molecular state and associated non-local Josephson effect. We demonstrate the peak structures for both mutual and self-inductance, and compute them in various frequency regimes. 

In an interesting common case the embedding circuit includes an oscillator, which can be used both to enhance hybridization and for state readout with two-tone spectroscopy. We derive and solve Lindblad equations for the conditions of two-tone spectroscopy to demonstrate the the readout of molecular states. If the readout oscillator can be chosen independently from the one used to enhance the hybridization we demonstrate that the states can be immediately identified from the oscillator response. However, in a more restricted setup where the same oscillator is used, the oscillator response manifests more resonant peaks indicating all transitions between states at different photon numbers.  
\end{abstract}

\maketitle

\section{\label{sec:Introduction}Introduction}
Hybrid superconducting devices are currently in focus of attention of condensed matter community. There are important applications of these devices in quantum sensing~\cite{jeanneretApplicationJosephsonEffect2009}, and they are implemented as platforms to realize Majorana~\cite{beenakkerSearchMajoranaFermions2013}, and superconducting~\cite{kjaergaardSuperconductingQubitsCurrent2020} qubits.
Low dissipation and disorder-robust phase coherence~\cite{bardeenTheorySuperconductivity1957, andersonTheoryDirtySuperconductors1959} make superconducting devices ideal for observing macroscopic quantum effects. Their natural integration into electric circuits also allows for efficient readout and manipulation of superconducting qubits, using the toolkit of circuit quantum electrodynamics~\cite{blaisCircuitQuantumElectrodynamics2021}.

\begin{figure}[b]
    \includegraphics[width=\linewidth]{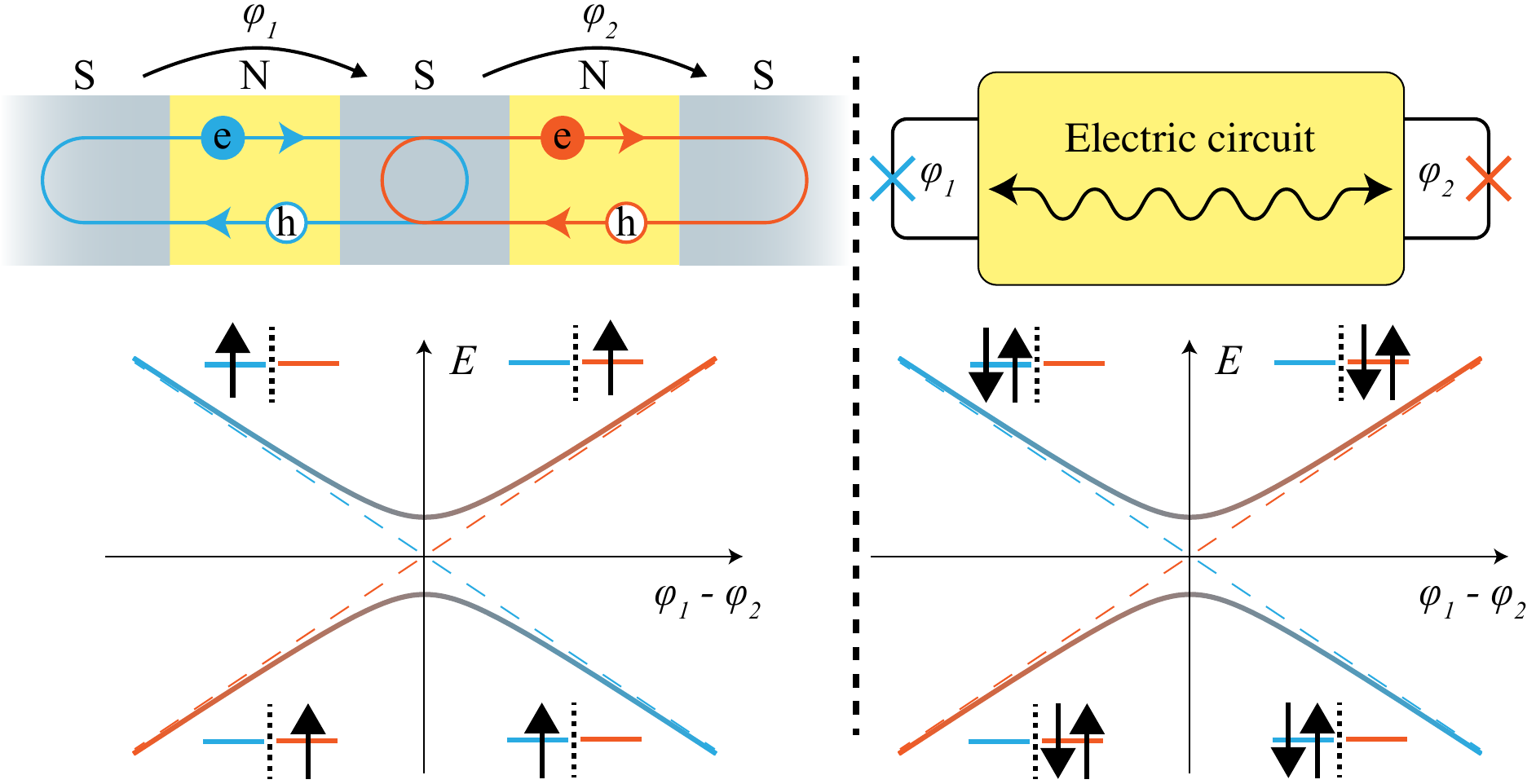}
    \caption{\label{fig:short-distance-long-distance} Left: Andreev molecule at short distances. The energies of quasiparticle states (with spin up) versus flux.  If the ABS energies at two different junctions (blue and orange) are aligned, weak electron transfer between the junctions hybridizes the states resulting in level repulsion and delocalized superpositions.	
	Right: Andreev molecule at long distances. The energies of excited singlet states versus flux. Since the junctions are separated by a large distance, the electron transfer is impossible. However, the junctions are coupled via a shared electric circuit. This also results in hybridization, level repulsion, and delocalized superpositions. }
\end{figure}

In a Josephson junction (JJ),  the interference of Andreev-reflected electrons and holes results in a set of discrete localized current-carrying Andreev Bound States (ABSs), a state appearing at each transmission channel and spin projection~\cite{andreevThermalConductivityIntermediate1964,beenakkerJosephsonCurrentSuperconducting1991,beenakkerUniversalLimitCriticalcurrent1991,furusakiCurrentcarryingStatesJosephson1991,furusakiJosephsonCurrentCarried1999}. Recently the setups where so-called "Andreev Molecules" (AM) can be realized have received ample theoretical~\cite{pilletNonlocalJosephsonEffect2019,pilletScatteringDescriptionAndreev2020,pilletJosephsonDiodeEffect2023,kornichFineEnergySplitting2019,kornichOverlappingAndreevStates2020, zsurkaNonlocalAndreevReflection2023,matute-canadasQuantumCircuitsMultiterminal2024, kurtossyHeteroatomicAndreevMolecule2024, zalomDoubleQuantumDot2024, kocsisStrongNonlocalTuning2024,hodtOnoffSwitchSign2023} and experimental~\cite{kurtossyAndreevMoleculeParallel2021,suAndreevMoleculesSemiconductor2017,haxellDemonstrationNonlocalJosephson2023,matsuoPhasedependentAndreevMolecules2023,matsuoObservationNonlocalJosephson2022, matsuoJosephsonDiodeEffect2023, scherublTransportSignaturesAndreev2019,jungerIntermediateStatesAndreev2023,matsuoPhaseEngineeringAnomalous2023} attention. In such setups, several Josephson junctions are brought sufficiently close to each other to enable  coherent electron and hole transfer between the junctions. This restricts the spacing between the junctions to a {\it short} length: typically, to the superconducting correlation length, either in the superconductor or in a non-superconducting material of the junctions (see e.g.~\cite{kornichFineEnergySplitting2019,kornichOverlappingAndreevStates2020} for detailed analysis).  The term AM refers to the situation of a relatively weak coupling between two ABS localized at different junctions.  The dependence of the ABS energy on the superconducting phases of the leads permits aligning these two energies.
In this case, the ABS are hybridized by even weak coupling, and the  superpositions formed are similar to the molecular orbitals in atomic physics, that is, delocalized over the junctions. 
 This is illustrated in the left side of Fig. \ref{fig:short-distance-long-distance}.
Generally, in an AM setup the superconducting current in each junction depends on the phase drops over both junctions. Thus it is possible to realize a non-local Josephson effect where the phase across one junction can be used to control the current across the other junction~\cite{matsuoObservationNonlocalJosephson2022,haxellDemonstrationNonlocalJosephson2023}.  Exotic single junction effects can also be realized by fixing a finite phase drop across one background junction to effectively break the time-reversal for the remaining junction. This has been used to observe the Josephson diode effect~\cite{matsuoJosephsonDiodeEffect2023}, and the anomalous Josephson effect in a $\phi$ Junction~\cite{matsuoPhaseEngineeringAnomalous2023}.
At stronger electron coupling the resulting spectrum differs substantially from that of independent junctions. Under certain conditions, this may result in a gap closing where one of the ABS is pushed into the continuum~\cite{matsuoPhasedependentAndreevMolecules2023, kornichOverlappingAndreevStates2020}. 

In this paper, we elaborate on the possibility to realize Andreev Molecules with the junctions separated by long distances such that no coherent electron transfer between them is ever possible. The single-quasiparticle states are thus not coupled and not hybridized. However, if two junctions are embedded in an electromagnetic environment, their {\it excited singlet states} can be coupled and hybridized by a photon exchange as illustrated in the right panel of~\ref{fig:short-distance-long-distance}. Such a state comprises two quasiparticles with opposite spin. Together with a ground singlet state, it constitutes an Andreev level qubit~\cite{zazunovAndreevLevelQubit2003, bretheauExcitingAndreevPairs2013, janvierCoherentManipulationAndreev2015} (ALQ)

In distinction from the short distance AM where the coupling is limited by superconducting correlation length, there is no distance restriction on the circuit-modulated coupling. It is only limited by the feasibility of manufacturing the circuits with low dissipation. 

In fact, the AM at long distance has been realized in a recent experiment by Cheung et al.~\cite{cheungPhotonmediatedLongrangeCoupling2024}  In the experiment, JJs etched in InAs nanowires with epitaxially grown aluminum shell were embedded in a superconducting loop that was inductively coupled to a NbTiN thin film superconducting resonator. This mediated the interaction between ALQs in the junctions. As a matter of fact, the junctions were separated by  \SI{6}{mm}. The hybridization of the ALQs has been confirmed via the observation of an avoided crossing by the pulsed two-tone spectroscopy ~\cite{cheungPhotonmediatedLongrangeCoupling2024}.

Another rather relevant recent experiment was performed by Pita-Vidal et al.~\cite{pita-vidalStrongTunableCoupling2024} and achieved long-range coupling between Andreev spin qubits~\cite{padurariuTheoreticalProposalSuperconducting2010,haysCoherentManipulationAndreev2021} by coupling spin-dependent supercurrents via an external circuit. Such setups are promising to achieve the hybridization of Andreev spin qubits at long distances. However, in the present paper we concentrate on ALQ.

We provide general understanding of the phenomenon and thoroughly discuss its manifestations in the context of experimental observation combining elements of qubit physics with specifics of ABS.
We establish  a model for several driven JJ embedded in a generic linear electric circuit and derive the associated Lindblad equations that describe the resulting spectrum as well as the dissipative dynamics. The specifics of the circuit are incorporated in a set of impedances. 
To emphasize the generality, we show which parameters in the Lindblad equation  may be derived just from classical circuit theory not invoking any microscopic model. 
 We identify and explore various regimes in the spectrum of the driven AM and investigate how the hybridization affects the characteristics of the steady state.  
 The important feature of AM is the non-local Josephson effect whereby the current in one junction depends on the phase drop on another one. In our configuration, it cannot be accessed via the measurement of critical current. However, it is manifested in a finite mutual inductance between the driven JJs. Both mutual and self-inductance peak near the resonance, and we thoroughly study the effect including its frequency dependence in different regimes.

 An oscillator can be a natural part of the embedding circuit, and is widely used in qubit setups both for tuning the couplings between the qubits and the actual qubit measurement by means of two-tone spectroscopy. We derive the effective Lindblad equation to describe the two-tone spectroscopy of an AM. In setups where only a single oscillator is used for both state readout and to enhance the ALQ coupling we show that the hybridization between the ALQs and that between the oscillator and the respective ALQs are necessarily of similar scale. This complicates the oscillator response due to a large feedback on the AM system from a single photon excitation. However, by using independent oscillators we show that the superpositions in the driven AM can be readily read out from from the oscillator response in the linear regime. The feedback on the AM system only becomes relevant for sufficiently strong driving. 

 While in the experiment~\cite{cheungPhotonmediatedLongrangeCoupling2024} both ALQ resonant frequencies were aligned to the oscillator as close as possible, we find advantageous the situation where a substantial frequency difference allows a good separation between the oscillator states and those of AM.

The paper is structured as follows. In Section~\ref{sec:Setup} we describe the general setup in use and give a simplified description based on circuit theory. We derive and discuss  the  Lindblad equation for AM starting from the microscopic Hamiltonian in Section~\ref{sec:Microscopic_model}, providing the details in Appendix A. We study the spectrum of AM in the same Section. We analyze the steady state of the driven AM  in Section~\ref{sec:Steady_state}.  Section~\ref{sec:Mutual_junction_inductance} is devoted to the evaluation of the mutual junction inductances in various frequency regimes. We incorporate the two tone setup into the Lindblad equation and study the AM dependent oscillator response in Section~\ref{sec:two_tone}, details are provided in Appendix~\ref{sec:appendixB}. Finally we conclude in Section~\ref{sec:Conclusion}.

\section{\label{sec:Setup}The setup and circuit theory analysis}

In this Section we describe the setup under consideration comprising two JJs embedded in a general linear electromagnetic environment, optionally comprising an oscillator. We perform circuit theory analysis approximating the JJs with linear circuit elements. While this is a crude approximation, we are able to correctly evaluate important quantities characterizing the AMS: energy splitting at the degeneracy point and the state dependent frequency shift of the oscillator that enables the AMS readout. We also establish the relevant scales for these quantities.

\begin{figure}
    \includegraphics[width=\linewidth]{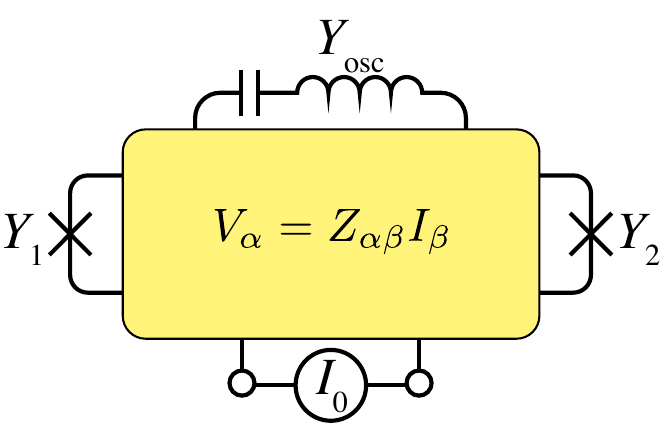}
    \caption{\label{fig:general-setup} Andreev molecule at long distance. Two JJs with admittances $Y_{1, 2}$ are embedded in a linear electric circuit characterized by the impedance matrix $Z_{\alpha \beta}$, where $\alpha$ and $\beta$ are used to label the ports. The circuit provides the coupling and hybridization between the excited states in the two junctions. 
    An oscillator (admittance $Y_{\rm osc}$) can also be included to the circuit.
    The system can be excited with an AC current source connected at port "0".}
\end{figure}

The setup is sketched in Fig \ref{fig:general-setup}.
The electromagnetic environment is a linear electric circuit defined by an impedance matrix $Z_{\alpha \beta}$ relating currents $I_\beta$ and voltages $V_\alpha$ across a set of ports at fixed frequency $\omega$
\begin{equation}
    V_\alpha(\omega) = \sum_\beta Z_{\alpha \beta}(\omega) I_\beta(\omega).
\end{equation}
An optional oscillator can be connected to port "3" where the current and voltage are related by the oscillator admittance 

\begin{equation}\label{eq:admittance_definition}
    I_{3}(\omega) = Y_{\text{osc}}(\omega) V_{3}(\omega),\quad \quad
    Y_{\text{osc}}(\omega)= \frac{1}{\frac{1}{-i\omega C} -i \omega L}.
\end{equation}
Equivalently the oscillator can be included to the electric environment with a corresponding change in $Z_{\alpha \beta}$ relating the remaining ports. 

Now we introduce an approximation replacing the JJs with linear circuit elements characterized by admittance $Y_\alpha(\omega)$, so
\begin{equation}
    I_\alpha(\omega) = Y_\alpha(\omega)V_\alpha(\omega).
\end{equation}
We are mostly interested in the part of the admittance describing resonant absorption at the ALQ frequency $2E_A$. In the vicinity of this frequency the JJ admittance is well approximated by a simple pole 
\begin{equation}\label{eq:JJ_admittance}
    Y(\omega) = i\bar{Y} \frac{2E_A}{\omega - 2E_A}, 
\end{equation}
where $E_A= \Delta \sqrt{1 - T \sin^2\left(\varphi/2\right)}$ is determined by the superconducting gap $\Delta$, junction phase drop $\varphi$, and transmission coefficient $T$. The prefactor $\bar{Y}$ has been computed in~\cite{kosFrequencydependentAdmittanceShort2013} for a generic model of a short single channel superconducting junction,
\begin{equation}
    \bar{Y}=
    \frac{G_Q \pi}{4} \frac{T^2(1-T)}{\left(1-T + \cot^{2}\left(\varphi/2\right)\right)^2},
\end{equation}
where $G_Q=e^2/\pi \hbar$ is the conductance quantum. 

We concentrate on the situation where two ALQs with resonant frequencies $2E_A^{(1),(2)} \equiv \omega_{1,2}$ are closely tuned to a common frequency $\Omega$, $\omega_1 \approx \omega_2 \approx \Omega$. To determine the change in resonant frequencies caused by the presence of the electromagnetic environment, we excite the setup with a current source $I_0$ as illustrated in Fig.
 \ref{fig:general-setup} and calculate the correspondent effective impedance 
\begin{equation}
    V_0(\omega) = z_{\text{eff}}(\omega) I_0(\omega).
\end{equation}
In general the effective impedance is given by
\begin{equation}\label{eq:zeff}
    z_{\text{eff}} = Z_{00} + \sum_{\alpha \beta \neq 0} Z_{0\alpha } \left(M^{-1}\right)_{\alpha \beta}Z_{\beta 0},
\end{equation}
where 
\begin{equation}
    M_{\alpha \beta} = Y^{-1}_{\alpha}\delta_{\alpha \beta} - Z_{\alpha \beta}
\end{equation}
 is defined for $\alpha, \beta \neq 0$ and the poles of $z_{\text{eff}}$ give the resonant frequencies of the system.

At frequencies close to $\Omega$ we can disregard the frequency dependence of $Z_{\alpha \beta}$, but keep the pole structure of $Y_{1, 2}$ (Eq.~\eqref{eq:JJ_admittance}). The poles of $z_{\rm eff}(\omega)$ are then given by the roots

\begin{equation}
    \det
    \begin{pmatrix}
        \frac{1}{i \bar{Y}_1} \frac{\omega - \omega_1}{\Omega} - Z_{11}  & Z_{12}\\
        Z_{21} & \frac{1}{i \bar{Y}_2} \frac{\omega - \omega_2}{\Omega} - Z_{22} 
    \end{pmatrix}
    =0
\end{equation}
The diagonal impedances $Z_{\alpha \alpha}$ give rise to a pole shift even in the absence of ALQ interaction ($Z_{12} = Z_{21}=0$): $ \omega_{\alpha} \rightarrow \omega_{\alpha}\left(1 + i \bar{Y}_{\alpha} Z_{\alpha \alpha}\right)$.
The imaginary part of the shift gives a finite decay rate 
\begin{equation}\label{eq:gamma}
    \gamma_\alpha = \omega_\alpha \bar{Y}_\alpha \Re\{Z_{\alpha \alpha}\}
\end{equation}
of the ALQs, and the real part determines an environment-induced energy shift. As we see in the following section the approximation in use give the wrong value of this shift, besides the shift is not observable since experimentally the only the renormalized frequency shift is accessible. Therefore we the real shift in a redefinition of $\omega_\alpha$.

In the presence of interaction between the ALQs, the poles of $z_{\text{eff}}(\omega)$ are at
\begin{equation}
    \bar{\omega}_{\pm} = \frac{\tilde{\omega}_1 + \tilde{\omega}_2}{2} \pm \sqrt{\left(\frac{\tilde{\omega}_1 - \tilde{\omega}_2}{2}\right)^2 - \Omega^2\bar{Y}_1 \bar{Y}_2 Z_{12} Z_{21} },
\end{equation}
where $\tilde{\omega}_\alpha\equiv\omega_\alpha+i\gamma_\alpha$.  If $\Im\{Z\} \gg \Re\{Z\}$ we can neglect the decay rates so the poles are close to the real axis,
\begin{equation}
    \bar{\omega}_{\pm} = \frac{\omega_1 + \omega_2}{2} \pm \sqrt{\left(\frac{\omega_1 - \omega_2}{2}\right)^2  + \lambda^2 },
\end{equation}
where
\begin{equation}\label{eq:lambda}
    \lambda \equiv \Omega \sqrt{\bar{Y}_1\bar{Y}_2} \frac{\Im \{Z_{12}\} +\Im\{Z_{21}\}}{2}.
\end{equation}
This reproduces the picture of an avoided crossing of hybridized molecular levels with a minimal splitting $\lambda$.

Let us now focus on frequencies close to the oscillator frequency $\omega_{\text{osc}} = 1/\sqrt{CL}$ where we approximate the oscillator admittance with
\begin{equation}
    Y_{\text{osc}}(\omega) = i \bar{Y}_{\text{osc}} \frac{\omega_{\text{osc}}}{\omega - \omega_{\text{osc}}}, \quad \quad 
    \bar{Y}_{\text{osc}} = \frac{1}{2} \sqrt{\frac{C}{L}}.
\end{equation}
Similarly (see Eq.~\eqref{eq:zeff}) we obtain the shift of the oscillator frequency $\delta \omega_{\text{osc}}$ in the presence of the ALQs
\begin{equation}\label{eq:oscillator_frequency_shift}
   \frac{\delta \omega_{\text{osc}} }{\omega_{\text{osc}}}=i \bar{Y}_{\text{osc}} \sum_{\alpha=1, 2} Z_{3\alpha} Y_\alpha(\omega_{\text{osc}})Z_{\alpha3}.
\end{equation}
Let us recall that the ALQs are qubits that can be in two states $\sigma_\alpha^z=\pm1$ and Eq.~\eqref{eq:JJ_admittance} only applies to the ground state, while the excited state admittance comes with opposite sign. Therefore the frequency shift Eq.~\eqref{eq:oscillator_frequency_shift} is state dependent 
\begin{equation}\label{eq:oscillator_state_dependent_shift}
    \frac{\delta \omega_{\text{osc}} }{\omega_{\text{osc}}}=- \bar{Y}_{\text{osc}} \sum_{\alpha=1, 2} Z_{3\alpha} \frac{\sigma^z_\alpha\bar{Y}_\alpha \omega_\alpha}{\omega_{\text{osc}} - \omega_\alpha} Z_{\alpha3},
\end{equation}
which enables the readout of the ALQ states \cite{blaisCircuitQuantumElectrodynamics2021}.

To obtain an order of magnitude estimate of the scales we first assume all circuit impedances are of the same order as the free space impedance $\sim \alpha/G_Q$, where $\alpha \approx 1/137$ is the fine structure constant. Since $\bar{Y}_{1, 2} \sim G_Q$
\begin{equation}
    \frac{\lambda}{\Omega} \sim \alpha,
\end{equation}
the interaction induced splitting is small compared to the ALQ energies. Similarly we obtain the formal estimate $\gamma_{1, 2} \sim \alpha \Omega$, however if $\Re\{Z\} \ll \Im\{Z\}$, $\gamma_{1, 2} \ll \lambda$, which makes it possible to resolve the avoided crossing.

To enhance $\lambda$, one increases impedance at $\omega=\Omega$, that can be readily done with an oscillator if $\omega_{\text{osc}} \approx \Omega$. Assuming $\bar{Y}_{\text{osc}} \sim G_Q/ \alpha$,
\begin{equation}
    \frac{\lambda}{\Omega} \sim \alpha \frac{\omega_{\text{osc}}}{|\Omega - \omega_{\text{osc}}|}.
\end{equation}
This enhancement saturates for $\lambda \sim |\Omega {-} \omega_{\text{osc}}|$ where the qubits are strongly hybridized with the oscillators (as was observed in \cite{cheungPhotonmediatedLongrangeCoupling2024}), so the maximum splitting $\lambda/\Omega \sim \sqrt{\alpha}$.

Similarly from Eq.~\eqref{eq:oscillator_state_dependent_shift} we estimate the state dependent shift of the oscillator energies
\begin{equation}
    \frac{\delta \omega_{\text{osc}} }{\omega_{\text{osc}}} \sim \alpha  \frac{\omega_{\text{osc}}}{{|\Omega-\omega_\text{osc}|}},
\end{equation}
which also saturates at $\delta \omega_{\text{osc}}/\omega_{\text{osc}} \sim \sqrt{\alpha}$. We see that $\lambda$ and $\delta \omega$ are of the same scale.

In these estimations we assumed $\bar{Y}_{1,2} \sim G_Q$, which is valid for $E_A \sim \Delta$. In fact most experiments are performed in a more convenient frequency region $\omega \ll \Delta$. The ABS energies are in this region provided $T\approx 1$ and $\varphi \approx \pi$, and can be estimated $E_A\sim \Delta\sqrt{1-T}$. This enhances the junction admittances by a big factor 
\begin{equation}
    \bar{Y} = \frac{G_Q \pi}{4} \left(\frac{\Delta}{E_A} \right)^2.
\end{equation}
This can be accounted for in previous estimations with the replacement $\alpha \rightarrow \alpha \left( \Delta/E_A\right)^2$. We conclude that the relative scales of $\lambda$ and $\delta \omega_{\text{osc}}$ are sufficiently large for their efficient experimental observation.

\section{\label{sec:Microscopic_model}Quantum model}

\begin{figure*}
    \includegraphics[width=\linewidth]{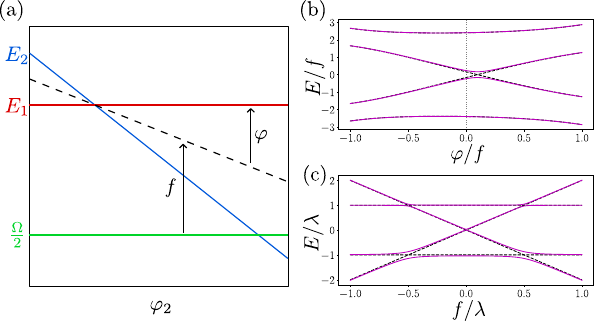}
    \caption{\label{fig:f_phi} (a) Convenient "coordinates" $f$ and $\varphi$. The red and blue lines 
    give the half-splitting of the first and second ALQ, while the green line is at half the oscillator frequency. The black dashed line gives the average qubit half-splitting $(E_1{+}E_2)/2$. All lines are plotted versus the superconducting phase difference across the second junction junction in the vicinity of the degeneracy point $E_1=E_2$.
    The "coordinates" $\varphi$ and $f$ are defined as shown. (b) The eigenvalues of $H_{{\rm eff}}$ in the large drive limit at $A_1 = 0.5 f$, $A_2 = 0.8f$, and $\theta = 0$ versus $\varphi$. The black dashed lines correspond to $\lambda=0$ while the magenta lines give the spectrum at $\lambda = 0.2 f$ where the degeneracy is lifted.  (c) The eigenvalues of $H_{{\rm eff}}$ in the small drive limit at $\varphi = 0$ and and $\theta=\pi{/}3$ versus $f$. The black dashed lines correspond to $A_1=A_2=0$, while the magenta lines correspond to $A_{1} = A_2 = 0.1\lambda$. Finite gaps linear in $A_{1}$ and $A_{2}$ open at $f = \pm\sqrt{\varphi^2 + \lambda^2/2}$. Note the asymmetry of positive and negative energies around the gaps while the limiting spectrum is mirror-symmetric. The degeneracy of $|ee\rangle$ and $|gg\rangle$ at $f=0$ is lifted with a much smaller gap quadratic in the drive. }
\end{figure*}

In this Section, we develop the quantum model: we provide a Lindblad equation that governs the  dissipative dynamics of an exited AM embedded in an arbitrary electric circuit and discuss its Hamiltonian part. While some results reproduce simple-minded derivation  of Sec. \ref{sec:Setup}, the full quantum description is obviously required to correctly describe dynamics of the associated ALQ: those are qubits rather than oscillators.  

We consider the system pictured in Fig \ref{fig:general-setup}, yet only concentrate  on two relevant states in  each JJ: ground and excited singlet forming the ALQs with energy splittings $2E^{(1)}_{A}(\varphi_1)$ and $2E^{(2)}_{A}(\varphi_2)$. A single ALQ interacting with the deviation of flux drop on the junction $\Phi(t)$ is described by the following Hamiltonian:
\begin{align}
\label{eq:Ham}
H = E \sigma_z - \hat{I} \Phi(t);\; \hat{I} = I_\parallel \sigma_z+ I_\perp \sigma_x.
\end{align} 
Here $\hat{I}$ is the operator of current in the junction and $I_\parallel = (2e/\hbar) \partial_\phi E$, $I_\perp = I_\parallel {\rm tan}\frac{\varphi}{2}$. 

Since we are interested in AM states, we need to choose the superconducting phases $\varphi_{1,2}$ such that $E_1 \approx E_2$, and excite the system. We do the latter by applying the oscillating flux drives to the junctions($\alpha=1,2$), $\Phi_{\alpha}(t) = \frac{1}{2} ( \Phi^{(d)}_\alpha e^{- i \Omega t} + \Phi^{(d)*}_\alpha e^{i \Omega t}) $, at a near-resonant frequency $\Omega \approx 2 E_1 \approx 2 E_2$.  We perform RWA transformation with respect to the frequency $\Omega$ to arrive at the Hamiltonian of two ALQ
\begin{align}
H &= (f + \varphi) \sigma_1^z + (f - \varphi) \sigma_2^z \nonumber \\
&-\sum_\alpha \frac{I_{\perp,\alpha}}{2} (\sigma^+_\alpha \Phi^{(d)}_\alpha + \sigma^-_\alpha \Phi^{(d)}_\alpha),
\end{align}
where we have introduced the convenient "coordinates" of dimension frequency: 
$\varphi \equiv (E_1-E_2)/2$ giving the distance from the degeneracy line in the space $(\varphi_{1},\varphi_{2})$ and $f\equiv (E_1+E_2 -\Omega)/2$ giving the drive frequency mismatch as illustrated in Fig \ref{fig:f_phi} (a). The scale of the resulting Hamiltonian is set by $f,\varphi \ll \Omega$. We have also neglected the terms $\propto I_\parallel$ describing the response to the low-frequency flux. Thereby we assume that the important circuit modes are at high frequency $\approx \Omega$. We will come back to these terms in Sec. \ref{sec:Mutual_junction_inductance}.

 At this stage, the Hamiltonians of the two ALQs are still uncoupled. To derive the coupling and dissipative dynamics induced by the environment, we treat the circuit as a collection of bosonic modes, express the operators of flux in terms of boson creation/annihilation operators, and derive the evolution equation for the density matrix of ALQ's in the second order of perturbation theory. This standard derivation is detailed in Appendix \ref{sec:appendix_A}. In addition, we find it convenient to rotate the ALQ with respect to corresponding $z$-axes to exclude the phase of the oscillating drives from uncoupled Hamiltonian.

With this, we obtain the following Lindblad master equation: 
\begin{equation}\label{eq:Lindblad}
    \partial_t \rho = -i\left[H_{{\rm eff}}, \rho \right] + \sum_{\alpha} 2 \gamma_{\alpha} \mathcal{D}(\sigma_\alpha^-, \rho, \sigma_\alpha^+),
\end{equation}
where
\begin{align} \label{eq:Heff}
    H_{{\rm eff}} &= (f + \varphi) \sigma_1^z + A_1 \sigma_1^x + (f - \varphi) \sigma_2^z +  A_2 \sigma_2^x \nonumber\\
    &+ i\lambda \left(e^{i \theta}\sigma_1^+ \sigma_2^- - e^{-i \theta} \sigma_1^- \sigma_2^+\right).
\end{align}
Here, $\mathcal{D}$ denotes the Lindblad form:
\begin{equation}
    \mathcal{D}(A, \rho, B) = A \rho B - \frac{1}{2}\left(BA \rho - \rho BA\right).
\end{equation}
for any operators $A,B$.
 The drives are rescaled in units of frequency $A_{\alpha} \equiv I_{\perp,\alpha}|\Phi^d_{\alpha}|$. The relative phase of the drives $\theta$  enters the coupling term $\propto \lambda$. As promised $\lambda$ is given by  Eq.~\eqref{eq:lambda} and the ALQ decay rates $2 \gamma_{\alpha}$ are given by the real pole shifts in circuit theory (see Eq.~\eqref{eq:gamma}). We note that both circuit and quantum theory also predict non-diagonal terms in Lindblad dissipation forms and induce decay correlations between the junctions. In the main text of the article we disregard these effects, assuming negligible real parts of transimpedances.

In the remainder of this Section we analyze the spectrum of $H_{{\rm eff}}$. It contains four eigenvalues which sum to zero because $H_{{\rm eff}}$ is traceless. In the limit $\lambda, A_{1,2} \to 0$, the states $|gg\rangle$ and $|ee\rangle$ where both ALQs are either in the ground or excited state have eigenvalues $\mp 2 f$. The states with only one excited ALQ, $|eg\rangle$ and $|ge\rangle$,have eigenvalues $\pm f$. Both $\lambda$ and drives $A_{1,2}$ result in hybridization of these states, and associated level repulsion around the degeneracy points.  

Let us consider simple limits of small and large driving, $\lambda \gg A_{1, 2}$ and $\lambda \ll A_{1, 2}$ respectively. These limits result in well-defined distinct parameter regimes near the avoided crossings where the hybridization is most efficient. 

In the large driving limit the spectrum at $\lambda=0$ reads 
\begin{equation}
    \label{eq:spectrum_zero_interaction}
E =\pm\xi_1\pm\xi_{2},
\end{equation} 
where
\begin{equation}
    \label{eq:xi}
    \xi_{1,2} =\sqrt{(f\pm\varphi)^2 + A_{1, 2}^2}
\end{equation}
are the eigenvalues of individual ALQs corresponding to hybridized ground and excited singlets. The states $|g'e'\rangle$ and $|e'g'\rangle$ are degenerate at $\xi_1=\xi_{2}$. This degeneracy is lifted with opening a gap $\propto \lambda$ manifesting an AM state, as illustrated in Fig. \ref{fig:f_phi} (b). 

In the opposite limit of small driving the spectrum at $A_{1, 2}=0$ reads $E=\{\pm 2f, \pm\sqrt{4\varphi^2+\lambda^2} \}$. The states $|ee\rangle$, $|gg\rangle$ remain unaffected by the coupling while $|eg\rangle$, $|ge\rangle$ hybridize forming AM superpositions. The effect of a finite drive amplitude is to open a gap linear in $A_{1}$ and $A_2$ around the degeneracy points $2 f = \pm\sqrt{4\varphi^2 + \lambda^2}$ mixing an AM state with either $|ee\rangle$ or $|gg\rangle$. Another degeneracy occurs at $f=0$ for  $|ee\rangle$ and $|gg\rangle$. This degeneracy is lifted in the second order in $A_{1,2}$  involving two-photon processes. The spectrum is illustrated in Fig. \ref{fig:f_phi} (c). 

In both limits, we observe the mirror symmetry in the spectrum: if $E$ is an eigenvalue, $-E$ is also an eigenvalue. This symmetry is not generic. In general, one proves $\sigma_1^y \sigma_2^y H(\theta) \sigma_1^y \sigma_2^y = -H(-\theta)$, so that the spectrum is mirrored upon changing the sign of $\theta$. In Fig \ref{fig:f_phi} (c) the symmetry breaks down around the degeneracy points $f = \pm\sqrt{4\varphi^2 + \lambda^2} /2$, where it is not generally possible to choose $\theta=0$ by a gauge transformation. 

The AM physics is most prominent and completely described by $H_{{\rm eff}}$ in the limit where $\gamma_{1}$ and $\gamma_2$ are small compared to the eigenvalues of $H_{{\rm eff}}$. In the remainder of the text we assume that this holds,  so we may use the eigenbasis of  $H_{{\rm eff}}$ and neglect the non-diagonal terms in the density matrix $\rho=\sum\rho_i\ket{i}\bra{i}$. The Lindblad equation Eq.~\eqref{eq:Lindblad} becomes a master equation
\begin{equation}\label{eq:master}
    \partial_t \rho_{i} = \sum_{j}\left(\Gamma_{i\leftarrow j} \rho_{j} - \Gamma_{j\leftarrow i} \rho_{ i}\right),
\end{equation}
where the transition rates $\Gamma_{i\leftarrow j}$ between the states are given by 
\begin{equation}\label{eq:rates}
    \Gamma_{i\leftarrow j}=2\sum_{\alpha}\gamma_{\alpha}\bra{i}\mathcal{D}(\sigma_\alpha^-, \ket{j} \bra{j}, \sigma_\alpha^+)\ket{i}.
\end{equation}
The stationary solution of these equations determines the steady state to be discussed in the next Section.

\section{\label{sec:Steady_state}Steady state signatures of AM}

\begin{figure*}
    \includegraphics[width=\linewidth]{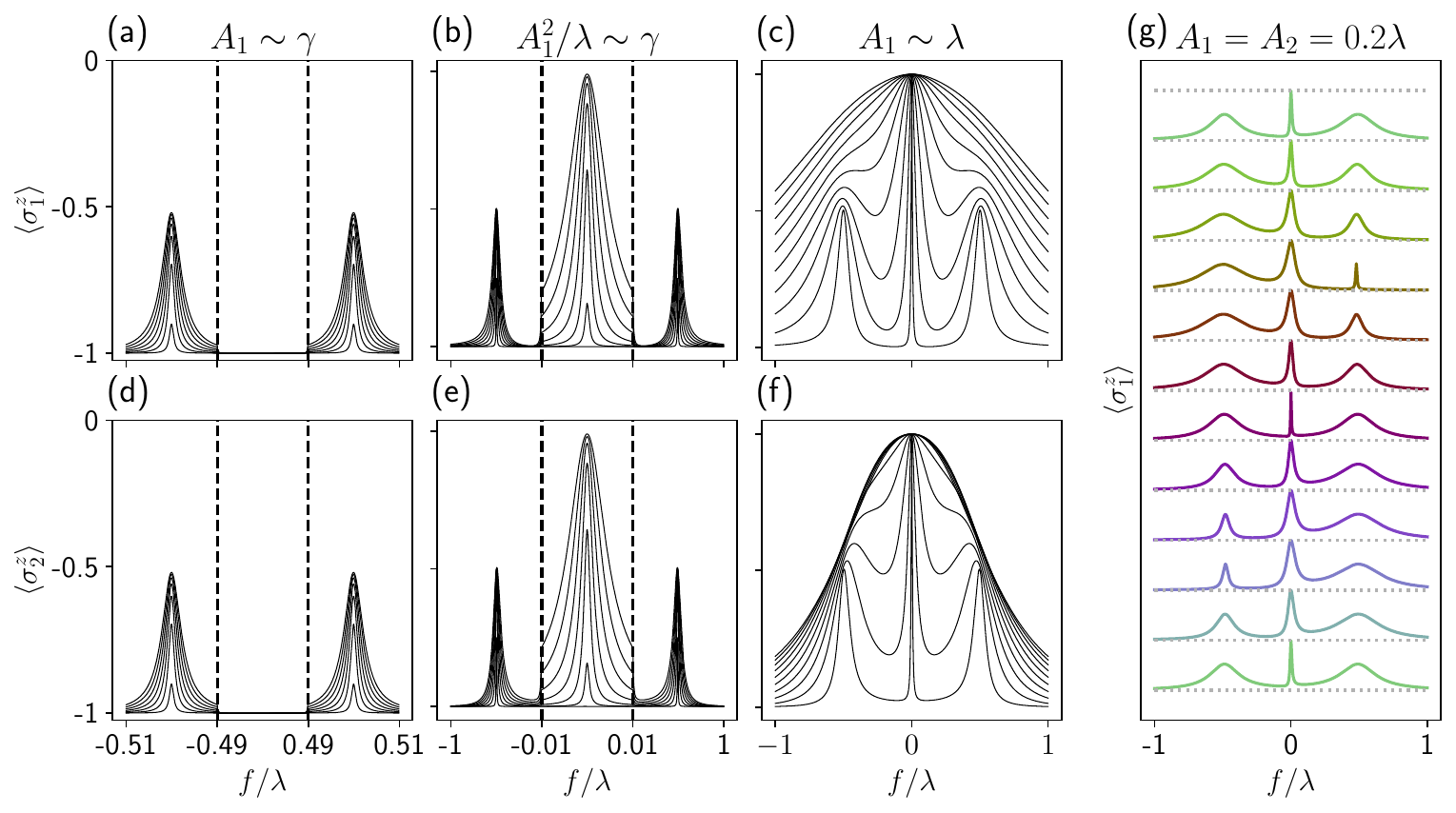}
    \caption{
	\label{fig:_interction_steadystate} 
	Degenerate ALQ. The excitation probabilities at $\varphi=0$ and $\theta=1.0$ versus $f$. Only the first ALQ is directly driven, $A_2=0$. The panes (a-c) and (d-f) show $\langle\sigma_1^z \rangle$ and $\langle\sigma_2^z \rangle$, respectively, and different $A_1$ increasing as the curves extends upwards. In the panes (a) and (d) $A_1$ increases linearly from curve to curve in the range $0.5 \gamma$ to $5 \gamma$. In the pane (b) and (e) $A_1$ increases linearly from curve to curve in the range  $0.16 \sqrt{\gamma\lambda}$ to $3.2 \sqrt{\gamma\lambda}$.
    In the panes (c) and (f), $A_1$ increases linearly from $0.1\lambda$ to $1.5 \lambda$. The vertical dashed lines and corresponding numbers indicate the changing horizontal scale. 
    The pane (g) shows  $\langle \sigma_1^z \rangle$ versus $f$ for $A_1{=}A_2{=}0.2\lambda$ and various values of their relative phase $\theta$. The different colored curves correspond to equally spaced $\theta$ in the range from $0.2$ to $ 0.2{+} 2\pi$. They are offset for clarity, so the distance between each dashed line corresponds to $\langle \sigma_1^z \rangle \in [-1, 0]$. In all plots, $\gamma_1{=}\gamma_2{=} 10^{-3} \lambda$.
	}
\end{figure*}

In this Section, we study the steady state solutions of the Lindblad equation (Eq.~\eqref{eq:Lindblad}) illustrating the experimentally observable signatures of the AM states. As the observable quantities, we choose the excitation probabilities $\langle \sigma_{1, 2}^z \rangle $ of the two ALQs. These quantities, in distinction from the state populations $\rho_i$,  can be directly measured by the commonly used techniques, for instance, by two-tone spectroscopy with a dispersively coupled oscillator \cite{blaisCircuitQuantumElectrodynamics2021}, see also Section \ref{sec:two_tone}. Moreover, $\langle \sigma_{1, 2}^z \rangle $ give the time-independent contributions to the JJ currents (see Section \ref{sec:Mutual_junction_inductance}). We note that this is not the only way to choose the observables: for instance, homodyne measurement at frequency $\Omega$ would characterize the oscillating currents in the junctions giving access to $\langle \sigma^+_{1,2}\rangle$, yet here we concentrate on the excitation probabilities.

The most obvious way to observe AM molecular states is to align the ALQ frequencies $\varphi=0$, and measure the excitation probabilities versus the drive frequency $f$. One expects to see resonances at $\pm \lambda$ signifying the split AM states, at least in the limit of small driving. Strong and generally asymmetric driving changes the resonant condition, and it is necessary to also change  $\varphi$ to reveal the hybridization signatures of the dressed ALQs. Accordingly, we separate the discussion into two parts: in subsection A we concentrate on degenerate qubits at $\varphi{=}0$, while in subsection B we fix $f$ to some value $f \gg \lambda$ and study the resonant features as function of $\varphi$.

\subsection{Degenerate qubits}
\begin{figure*}
    \includegraphics[width=\linewidth]{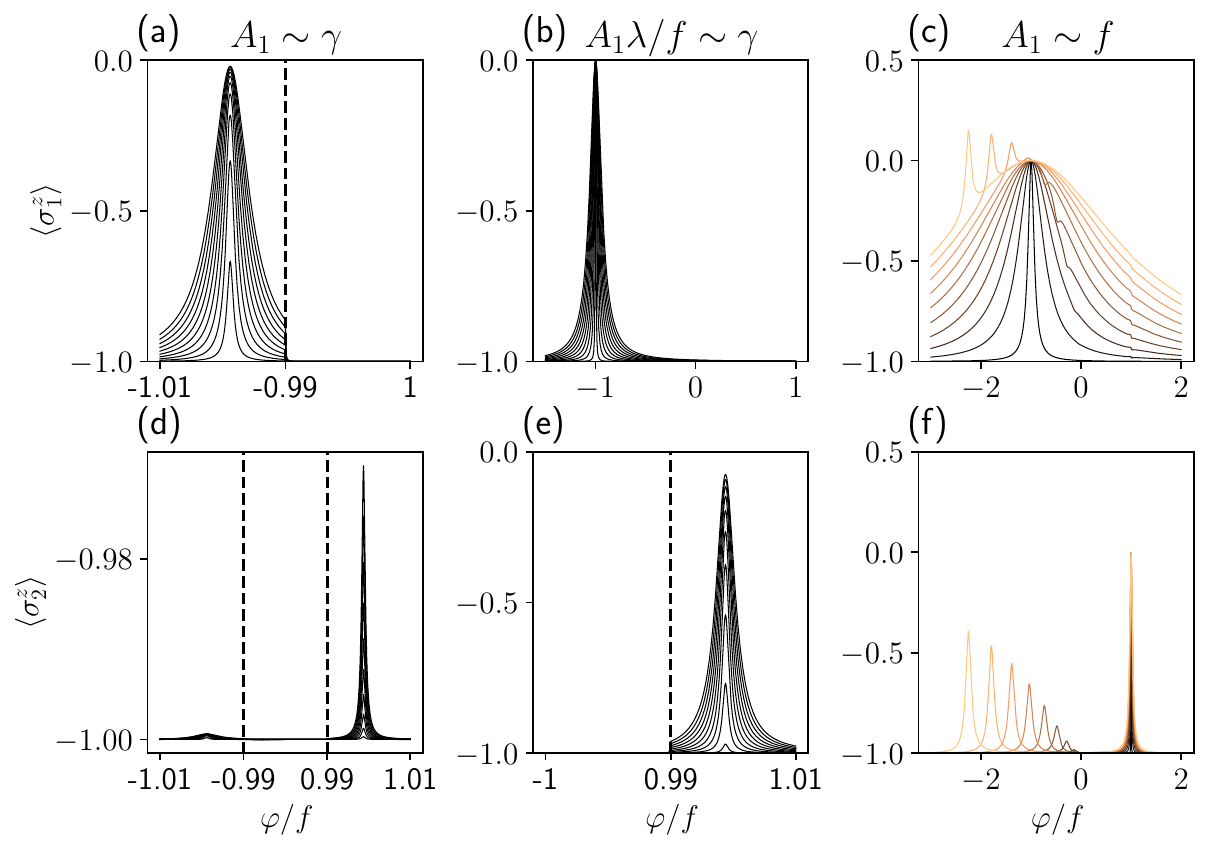}
    \caption{\label{fig:small_interction_steadystate} Tuning resonances with flux. The ALQ excitation probabilities for a fixed $f{=}10 \lambda$ versus $\varphi$. Only the first ALQ is driven directly, $A_2 =0$, and $\theta=1.0$. The panes(a-c) and (d-f) show $\langle\sigma_1^z \rangle$ and $\langle\sigma_2^z \rangle$ respectively for varying $A_1$. In the panes (a) and (d), $A_1$ increases linearly in the range from $0.5 \gamma$ to $5 \gamma$ from lower to upper curves. In the panes  (b) and (e), $A_1$ increases linearly from $0.5 {\gamma (f/\lambda)} $ to $10 {\gamma (f/\lambda)}$.
    In the panes (c) and (f), $A_1$ increases linearly from $0.1f$ to $3 f$ from the darkest to the lightest curve. For all plots, $\gamma_1{=}\gamma_2{=}10^{-3}f$. 
	The horizontal scale in (a), (b), (d), (e) is not uniform as indicated by the vertical dashed lines and corresponding numbers.}
\end{figure*}

We illustrate the frequency dependence of excitation probabilities in the degenerate case $\varphi{=}0$ in panes (a) through (f) of Fig. \ref{fig:_interction_steadystate}. Those  have been computed by solving the steady state Lindblad equation numerically with the QuTiP Python package \cite{johanssonQuTiPOpensourcePython2012} (See~\cite{samuelsenCodeFiguresData2025} for code to reproduce the figures in this article). We set decay rates to  $\gamma_1{=}\gamma_2 {\equiv} \gamma{=} 10^{-3} \lambda$. To make the presence of the delocalized AM states obvious, we apply the drive to the first qubit only, $A_2=0$, and vary the drive amplitude $A_1$.

We identity three different regimes and start with the regime of the relatively small drive, $A_1 \simeq \gamma$ (panes (a) and (d)).
There are resonant peaks at $f{=}{\pm}\lambda/2$ that correspond to excitation of one one of the AM superposition states, either $|+\rangle$ or $|-\rangle$ from the ground state. The AM states are equal-weight superpositions of the states $\ket{e g}$ and $\ket{g e}$,
$\ket{\pm} = (\ket{e g} \mp \ket{g e})/\sqrt{2}$. The corresponding spectrum is exemplified in Fig \ref{fig:f_phi}(c). We observe almost identical $\langle \sigma_z \rangle$ in both qubits, this manifests hybridized states. The heights of the peaks are proportional to $A_1^2/\gamma$ for small $A_1 \ll \gamma$ and saturate at $\langle \sigma_1^z\rangle = \langle\sigma_2^z\rangle= {-} 1{/}2$ if $A_1$ is several times bigger than $\gamma$. Such saturation level manifests the hybridization: indeed, at saturation  two states, $|g g\rangle$ and one of $\ket{\pm}$
are equally populated, and, since  $\langle g g| \sigma^z_{1, 2}  |g g\rangle = -1$, the observed excitation probabilities prove  $\langle \pm| \sigma^z_{1, 2}  |\pm\rangle =0$. The width of the peaks $\simeq \gamma$ at $A_1 \ll \gamma$
and $\simeq A_1$ at $A_1 \gtrsim \gamma$, the latter manifests the splitting $\simeq A_1$ of drive-hybridized states.

A less expected resonant feature becomes visible and dominates at larger drives $A_1^2\simeq\lambda\gamma$ (panes (b) and (e)  of Fig. \ref{fig:_interction_steadystate}). It emerges at $f=0$ due to a resonant two-photon absorption that sends the system to $|ee\rangle$ from the ground state $|gg\rangle$. The height of this  peak $\simeq A_1^4/\lambda^2\gamma^2$ at $A_1^2 \ll \lambda\gamma$. If  $A_1^2 \gtrsim /\lambda\gamma$, the peak saturates  at $\langle \sigma_1^z\rangle = \langle\sigma_2^z\rangle = 0$ corresponding to equal populations of $|gg\rangle$ ($\langle \sigma^z_{1,2} \rangle =-1$) and $|ee\rangle$ ($\langle \sigma_z \rangle =1$). The width of the peak correspondingly increases from $\simeq \gamma$ to  $A_1^2 /\lambda$, the latter agrees with the splitting $\simeq A_1^2 /\lambda$
of the drive-hybridized states.

In the regime of the larger driving,  $A_{1}\gtrsim \lambda$ ((c) and (f) of Fig. \ref{fig:_interction_steadystate}), all three widening peaks merge to one at $A_{1}{\simeq} \lambda$. At these values of the drive amplitude, we finally see the difference between the ALQ's: the peak in the non-excited ALQ saturates at  the width $\simeq \lambda$, while the width of the peak of the excited ALQ keeps growing being ${\simeq}A_1$. In this limit, the ALQs are tuned out of resonance by the drive, and the first ALQ is driven directly, while the saturation of the  peak in non-driven ALQ can be explained as a saturation of the coupling-mediated driving. To see this, let us substitute $\langle \sigma_1^{\pm} \rangle = 1{/}2$ to the coupling part of  the Hamiltonian Eq.~\ref{eq:Heff}: this gives rise to an effective driving of the second ALQ with amplitude $A^{\rm eff}_2 \sim \lambda/{2}$.

In Fig. \ref{fig:_interction_steadystate} (g) we demonstrate an interference effect of two drives that manifests itself in the dependence of the excitation probabilities on the relative phase of the drives $\theta$. As seen from Eq.~\ref{eq:Heff}, this phase affects the non-diagonal elements of the total driving amplitudes between the states $|gg\rangle$ and $|\pm \rangle$ as well as between the states $|ee\rangle$ and $|\pm \rangle$. To maximize the interference, we choose $A_1=A_2$. We also choose sufficiently large drive to saturate all three peaks. The modulation of the non-diagonal elements is then seen as the modulation of the peak widths. Changing $\theta$ in the interval $(0, 2\pi$), we observe that the peaks can be tuned to almost zero width. Another manifestation of this interference effect is that the magnitude of the avoided crossings at positive and negative eigenvalues (see (c) Fig. \ref{fig:f_phi}) can be tuned by changing $\theta$.

\subsection{Tuning resonances with the flux}
The AM states can  also be detected far from degeneracy. To demonstrate this, we fix  
the drive frequency $f \gg\lambda$ ($f=10\lambda$ for in the plots). Changing the flux $\varphi$, we can bring the ground state in resonance with $|eg\rangle$ at $\varphi = - f$ and with $|ge\rangle$ at $\varphi = f$. We intentionally choose to drive only the first ALQ, $A_2=0$. 
In the panes (a-f) of Fig. \ref{fig:small_interction_steadystate} we plot the excitation probabilities versus $\varphi$ at various $A_1$. The decay rates are set to 
$\gamma_1=\gamma_2\equiv\gamma=10^{-3}f$

In the weak drive regime $A_1 \simeq \gamma$, the driven ALQ (Fig. \ref{fig:small_interction_steadystate}  (a)) exhibits a standard peak at $\varphi {=} {-} f$ that saturates at $\langle \sigma_1^z \rangle=0$ for $A_1 \gtrsim \gamma$ with width $\simeq {\rm max}(\gamma, A_1)$, as if the second ALQ is not present at all. As for the non-driven ALQ ((Fig. \ref{fig:small_interction_steadystate}  (a)),
the residual AM hybridization gives a very small peak 
$\simeq (\lambda/f)^{-4}$ around this value of flux. 
A much larger peak is observed at $\varphi = f$ where the non-driven ALQ is at resonance with the drive. This is due to the coupling-mediated driving mentioned above. Its magnitude can be estimated as $A^{\rm eff}_2 \simeq\lambda \langle \sigma_1^\pm \rangle \sim \lambda A_1 / f$ in the saturation regime, corresponding to the peak height $\simeq (A_1/\gamma)^2 (\lambda/f)^2$. The width of this peak remains $\simeq \gamma$.

At further increase of $A_1$, the excitation probability peak saturates when $A_2^{\rm eff}$ reaches $\simeq \gamma$, that is, at $A_1 \simeq \gamma (\lambda/f)$ (see Fig. \ref{fig:small_interction_steadystate} (e)).

A more complex picture emerges for even larger drive amplitudes $A_1\gtrsim f$ where the hybridization becomes important near the crossing points of the dressed states. In the limit of large drive amplitudes, the crossing points require $\xi_1=\xi_2$ (see Eq.~\eqref{eq:spectrum_zero_interaction}), this gives the degeneracy of $|e' g\rangle$ and $|g' e\rangle$ at $\varphi=\varphi_c\equiv - A_1^2/4f$. 
Indeed, we see in Fig. \ref{fig:small_interction_steadystate} (c) narrow (width $\simeq \lambda$)
peculiarities at the flux value $\varphi_c$, at the background of the much wider (width $\simeq f$) standard peak. If 
$\varphi_c > - f$ ($A_1 < 4 f$), these peculiarities are rather small dips. At $\varphi_c < - f$, they become pronounced peaks of increasing magnitude. In Fig. \ref{fig:small_interction_steadystate} (f) we see the peaks of similar width at the same positions $\varphi=\varphi_c$: the excited state of the second ALQ is populated signifying AM physics. At $\varphi \approx f$ we still observe the saturated excitation peak driven by the coupling-induced $A_2^{\rm eff} \simeq \lambda$ and thus being of comparable width to the hybridization peaks.
Thus we reveal the pronounced signature of AM in the large driving limit.

To understand the details of this signature, we solve the master equation exactly at the crossing point in the limit where one of the drives is large compared to the interaction $\lambda \ll A_{1, 2}$ (see Appendix \ref{sec:steady_state_interaction}). In the general case where $A_2 \ne 0$, the crossing point is at $\varphi = \varphi_c \equiv (A_2^2-A_1^2)/4f$, and at this flux the excitation probability of the first ALQ reads: 

\begin{equation}\label{eq:peak_z1}
    \langle \sigma_1^z \rangle =  -(f+\varphi_c) \frac { f + \varphi_c (g_1 - g_2)}{ f^2 +\varphi_c^2 + \frac{1}{2} g_2 A_1^2 + \frac{1}{2}g_1A_2^2 },
\end{equation}
where $g_{1,2} \equiv \gamma_{1,2}/(\gamma_1+\gamma_2)$. 
In the case of Fig. \ref{fig:small_interction_steadystate} ($A_2=0$ and $g_1=g_2=1/2$),
\begin{equation}
    \langle \sigma_1^z \rangle =  -(f+\varphi_c) \frac { f }{f^2 + \varphi_c^2 + \frac{1}{4}A_1^2}.
\end{equation}
This changes sign as the crossing happens at $\varphi_c = - f$, or equivalently $A_1 = 2f$ as seen in Fig. \ref{fig:small_interction_steadystate} (c). 

Let us next consider very different ALQ decay rates and concentrate on the long-lived qubit, so that $\gamma_2\gg\gamma_1$, $g_2\to 1$, and $g_1 \to 0$. In this limit, 
\begin{equation}
    \langle \sigma_1^z \rangle \rightarrow  -\frac{f^2 - \varphi_c^2}{(f - \varphi_c)^2 + \frac{1}{2}A_{2}^2}.
\end{equation}
For large  $|\varphi_c| \gg f$ the value of $\langle \sigma_1^z \rangle $ reaches maximum value of 1 at the degeneracy point, indicating an almost pure state.

To understand the result, we note that for $\varphi_c \gg f $, which implies $A_{1,2} \gg f$, both ALQ's are out of the resonance and hardly dressed. In the absence of coupling, the system is in the ground state  $\ket{gg}$. If coupling is taken into account, the states $\ket{ee}$ and $\ket{gg}$ are degenerate. They are mixed with an effective gap $\tilde{\lambda} \simeq \lambda (f/\varphi_c)$ which is reduced in comparison with $\lambda$ because the dressing is small. At the degeneracy point, this hybridization causes equal populations of the degenerate states. 
 The fast decay rate of the second ALQ guarantees that $\ket{e e}$ only decays to $\ket{e g}$, and the system gets trapped in this state. A much slower decay of the first ALQ brings it back to $\ket{gg}$.
We note that this effective flipping of the second ALQ is limited to a narrow window of flux around $\varphi_c$, the widths being determined by the effective overlap $\tilde{\lambda}$ (see Appendix \ref{sec:steady_state_interaction} for the details).

\section{\label{sec:Mutual_junction_inductance}Inverse inductance as a signature of AM}
An important property of the JJ is to carry a finite DC supercurrent. In an AM state one expects non-local Josephson effect: the current in one JJ depends on the superconductive phase drop at another one.
One cannot measure this current immediately, and the measurements of switching current implemented in \cite{haxellDemonstrationNonlocalJosephson2023} for short AM are not suitable for our setup: it involves a full sweep of the phase and in our case the hybridization is restricted to a narrow window of superconducting phase difference. The valuable alternative is to measure the current response on a small change of flux, that is, inverse inductance. Throughout the Section, we use $K$ to denote inverse inductance.

The response of currents in our ALQs on a small flux amplitudes oscillating at frequency $\omega$ thus reads
\begin{equation}\label{eq:inductance}
    \langle \hat{I}_\alpha(\omega) \rangle = \sum_\beta K_{\alpha \beta}(\omega) \delta \Phi_\beta(\omega).
\end{equation}
This gives the oscillating current in junction $\alpha$ in response to an applied flux drive at  junction $\beta$.

The ALQ contribution to the inverse inductance of the whole circuit may be small, and a common
approach to measure it is to embed a low frequency oscillator in the circuit. One would then  measure the ALQ dependent  shift of the oscillator frequency, which is given by the low frequency analogue of Eq.~\eqref{eq:oscillator_frequency_shift} 
\begin{equation}
    \delta\omega_{\text{osc}} = -\bar{Y}_{\text{osc}}
    \begin{pmatrix}
        Z_{31} & Z_{32}
    \end{pmatrix}
    \begin{pmatrix}
        K_{11} & K_{12}\\
        K_{21} & K_{22}
    \end{pmatrix}
    \begin{pmatrix}
        Z_{13} \\
        Z_{23}
    \end{pmatrix},
\end{equation}
where $3$ is the port of the oscillator and $K$ is measured at the oscillator frequency. The ALQ part of the response can be singled out by taking into account its flux and drive dependence.

In the following we restrict the analysis to low frequencies $\omega \ll E_{A}$.
Even under this condition the matrix $K$ has a distinct frequency dependence arising from  the ALQ dynamics. The actual frequency at which it is measured is determined by the design of the low-frequency oscillator. This is why we need to explore $K(\omega)$ at various frequencies not restricting ourselves to the limit of zero $\omega$.

To compute $K$, we include the low-frequency flux terms $-\sum_\alpha  {I}_{\parallel,\alpha} \sigma^z_\alpha \delta \Phi_\alpha(t)$ in the original system Hamiltonian and concentrate on the current change due to these terms, that is also proportional to ${I}_{\parallel,\alpha}$.  We stress that within our model $K_{\alpha\beta}$ is always proportional to 
${I}_{\parallel,\alpha} {I}_{\parallel,\beta}$, so it makes sense to introduce the normalized inverse conductance matrix
$\tilde{K}_{\alpha\beta} = K_{\alpha\beta}/{I}_{\parallel,\alpha} {I}_{\parallel,\beta}$ that has dimension of inverse energy.

Our treatment involves two distant frequency scales: the low frequency ALQ decay rates $\gamma_{\alpha}$, and the higher frequency Hamiltonian level spacing $\Delta E$, both much smaller than $E_A$. 
Large separation between these scales naturally gives rise to three distinct frequency regimes: 
the zero-frequency limit $\omega \ll \gamma_{\alpha}$, the adiabatic regime $\gamma_{\alpha} \ll \omega \ll \Delta E$, and the resonant regime $\omega \sim \Delta E$. We discuss these regimes in separate subsections.

\subsection{Paired extrema feature}
It is important to keep in mind that we are working with resonantly excited ALQ and therefore the inverse inductance is much larger than that in the ground state and exhibits a specific resonant feature of paired extrema.
Let us see this from the analysis of a single ALQ.

The point is that the DC currents are opposite in the states $|g\rangle$ and $|e\rangle$. Since the populations of these states are changing essentially near the resonance, the shift of the resonant frequency by flux produces an enhanced response. Defining $\epsilon \equiv E_A - \Omega/2$ we derive
\begin{align}
K = I_\parallel \partial_{\Phi} \langle \sigma^z \rangle =  
I_\parallel \partial_\epsilon \langle \sigma^z \rangle \partial_{\Phi} \epsilon &=  I_\parallel^2 \partial_\epsilon \langle \sigma^z \rangle  \nonumber \\
&= -I_\parallel^2 \frac{\epsilon  A^2}{ \left(\epsilon^2 +\frac{1}{2}A^2\right)^2}
\end{align}  
where we substituted the expression  $\langle \sigma^z \rangle=- \epsilon^2/(\epsilon^2 + A^2/2)$ valid for sufficiently large drive $A \gg \gamma$.

If plotted versus either frequency or flux, $K$ exhibits a typical feature of two paired extrema with separation on the order of the resonant width. 
The maximum value of $K$ is estimated as 
$I_\parallel^2/A \simeq I_\parallel/\Phi_0 (E_A/A)$, $\Phi_0$ being the flux quantum.
In the ground state, $K = \partial^2_\Phi E_A \simeq I_\parallel/\Phi_0$. Therefore, the resonant contribution to $K$ dominates if $A \ll E_A$. Note that the maximum absolute value of the inverse inductance is inversely proportional to the  width of the excitation peak.

\subsection{Zero-frequency limit}

\begin{figure*}
    \centering
    \includegraphics[width=\textwidth]{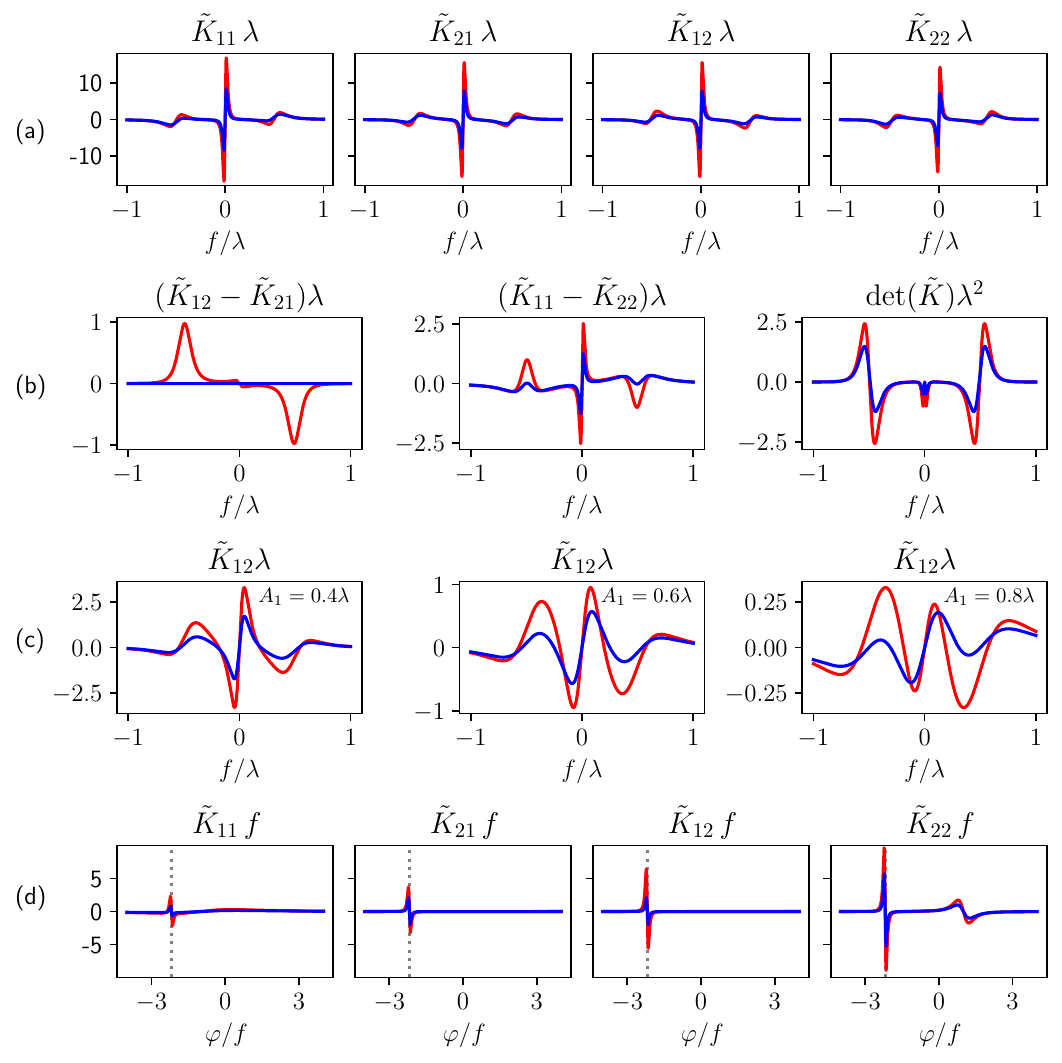}
    \caption{\label{fig:inductance_frequency_independent}  The normalized inverse inductance matrix in the zero-frequency limit (shown in red) and adiabatic regime (shown in blue). (a) The four components of $\tilde{K}$ versus $f$ for degenerate qubits. The settings are the same as in Fig. \ref{fig:_interction_steadystate} at $A_1=0.2 \lambda$. 
	All four components are approximately the  owing to strong hybridization.  (b) The differences of the components of $\tilde{K}$ at the same settings. Left pane: anti-symmetric part. Middle pane: the difference of diagonal elements. Right pane: the determinant of $\tilde{K}$. (c) The drive dependence of $\tilde{K}_{12}$ at the same other settings. The drive amplitude $A_1$ is given inside each pane. (d) Flux-tuned resonances. The four components of $\tilde{K}$ versus $\varphi$. The settings correspond to Fig. \ref{fig:small_interction_steadystate} at $A_1 = 3 f$ and $A_2 = 0.5 f$. 
	The dashed vertical line indicates the position of the degeneracy point $\varphi=\varphi_c$.}
\end{figure*}
In the zero-frequency limit $\omega \ll \gamma_{1,2}$ the system is at equilibrium with the time-dependent flux so the time-dependent current response is proportional to the derivative of the steady state current with respect to the flux, this defines the components of $K$, 
\begin{equation}\label{eq:dc}
    K_{\alpha \beta}(\omega \ll \gamma_{\alpha}) = \partial_{\Phi_\beta} \langle \hat{I}_\alpha \rangle = \sum_{i}\partial_{\Phi_\beta} \rho_{i}\partial_{\Phi_\alpha} E_i,
\end{equation}
where $i$ labels the eigenstates of $H_{\rm eff}$ and we made use of $\hat{I} {=}\partial_\Phi H$. 
We use the above formulas and the numerical solution of Lindblad equation to illustrate the peculiarities of $K$. 
In Fig. \ref{fig:inductance_frequency_independent} (a)  we concentrate on degenerate qubits, $\varphi=0$, taking the parameters corresponding to Fig. \ref{fig:_interction_steadystate} and choosing an intermediate drive $A_1=0.2 \lambda$ at which all three peaks at $f=\pm \lambda/2,0$ are saturated but still do not overlap. Zero-frequency results for all components of $K$ are plotted in red.
We observe a paired extrema feature for each resonant peak.
 As seen from the previous subsection, the natural units for $K_{\alpha \beta}$ are $I_{\parallel,\alpha} I_{\parallel,\beta}/\lambda$. However, as noted, the actual value of $K$ is determined by the corresponding peak width. It is higher for the two-photon peak since it is more narrow. It is seen from the plots that for these settings all the components of $\tilde{K}$ are approximately equal, despite the fact that the driving is not symmetric. This manifests strong hybridization of AM states. (If $I_{\parallel,1} \ne I_{\parallel,2}$ the components of non-normalized $K$ are {\it not} the same).

Let us concentrate on the differences between components 
(Fig. \ref{fig:inductance_frequency_independent} (b)). In the left pane, we plot anti-symmetric part of the mutual inverse inductance. This part should be zero in thermodynamic equilibrium and when time-reversibility holds, by virtue of Onsager relations. For instance, this holds in usual linear circuit theory. It is not zero in our case because the ALQ system is driven. We see that the anti-symmetric part peaks or dips at the positions of the AM resonant peaks and is odd in $f$ for these settings. In the middle pane, we plot the difference of the diagonal elements. We observe the same features: odd in $f$, peak/dip at $f \mp \lambda/2$. In addition, we see a pronounced paired extrema feature at $f=0$. 
In the right pane, we plot the determinant of $\tilde{K}$ that would also be 0 if all elements of the matrix are the same. 
The determinant is even in $f$ and exhibit paired extrema features at $f \mp \lambda/2$. It is worth noting that commonly the determinant of a matrix response function is always positive, this can be related to stability condition of a stationary point and/or positivity of the dissipation. We see that in our case the determinant changes sign, and is negative in a large interval of $f$. This should not be surprising since the ALQ system is driven. Depending on the properties of the embedding circuit, this may lead to electric instabilities in the setup.

Let us address flux-tuned resonances.
In Fig. \ref{fig:inductance_frequency_independent} (c), we plot the components of $K$ at the settings similar to those
of Fig. \ref{fig:small_interction_steadystate}, (c) and (f), for $A_1=3 f$, $A_2=0.5$.
(exact settings are specified in the figure caption). We see the correspondence between the peaks in excitation probabilities, for the first and second ALQ, and the paired extrema features in the diagonal elements of $\tilde{K}$. 
As to non-diagonal elements, they manifest only the peak at $\varphi=\varphi_c$. This is expected since only this peak is related to the hybridization of the dressed ALQ states and thus provides a mutual inverse inductance.

In Fig. \ref{fig:inductance_frequency_independent} (d) we demonstrate the dependence of a non-diagonal element of $\tilde{K}$ on the drive amplitude. The other settings are the same as for degenerate qubits, Fig \ref{fig:inductance_frequency_independent} (a).  This clearly shows that all the features  become broader upon increasing the drive and their magnitude decreases proportionally.

\begin{figure}
    \centering   
    \includegraphics[width=\linewidth]{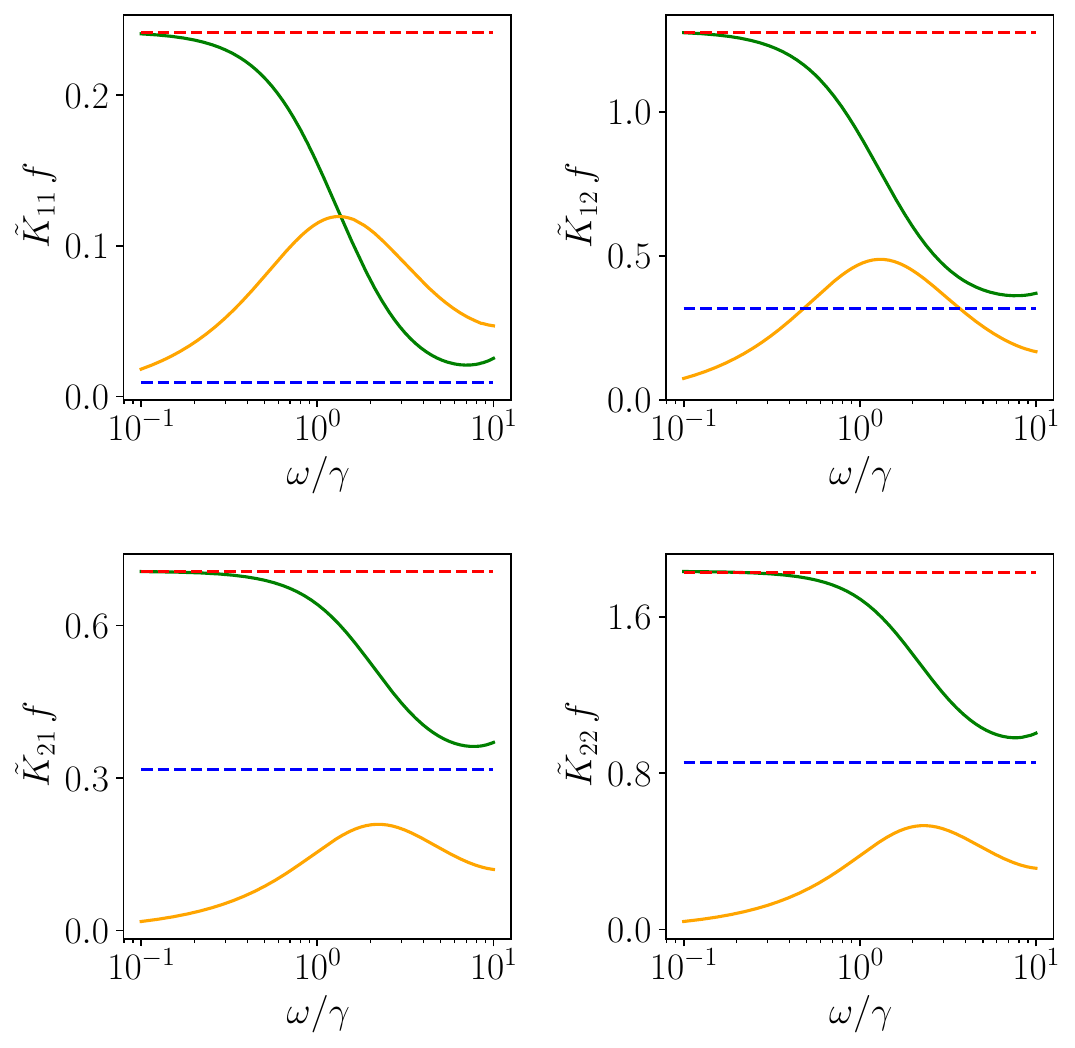}
    \caption{\label{fig:inductance_frequency_dependent} Real and imaginary parts of the inverse inductance matrix components versus frequency at the scale $\omega \sim  \gamma \equiv \gamma_1 {=}\gamma_2$. We illustrate the transition between zero-frequency and adiabatic regimes. The setting are identical to that of Fig. \ref{fig:inductance_frequency_independent} (d) at $\varphi = -2.4f$ except $\gamma =10^{-2} f$, $\lambda=0.2f$.
	The real and imaginary parts are given in green and orange color respectively. The dashed lines give the real part of the components in zero-frequency (red) and adiabatic (blue) limits.}
\end{figure}

\begin{figure*}
    \centering   
    \includegraphics[width=\textwidth]{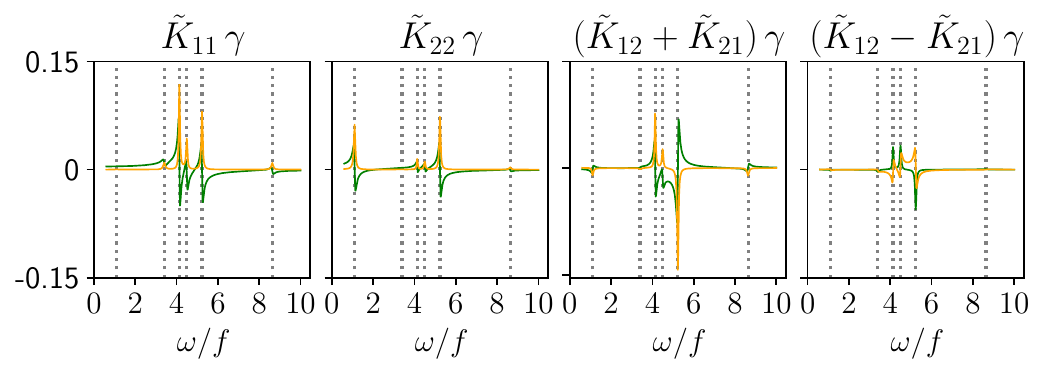}
    \caption{\label{fig:inductance_resonant} The frequency dependence of the normalized inverse inductance matrix in the resonant regime. The settings are: $A_1=2 f$, $A_2=f$, $\theta=1.0$, $\varphi= -f$, $\lambda =f$, and $\gamma_1=\gamma_2= 0.02 f$. At these settings, the eigenvalues are $E_1 =-4.50$,  $E_2=-0.37$, $E_3=0.73$, $E_4=4.13$. The dotted lines  indicate the positions of the resonances at eigenvalue differences.
	From left to the right, they correspond to transitions
	$ 3 \leftrightarrow 2$, $3 \leftrightarrow 4$, $1\leftrightarrow 2$, 
	$2 \leftrightarrow 4$, $1 \leftrightarrow 3$, $1 \leftrightarrow 4$,
	so all possible resonances are visible. 
	We plot the symmetric part of $\tilde{K}$ in (a), (b), (c), and the anti-symmetric part in (d).
	Real and imaginary parts are given in green and orange color respectively. 
	}
\end{figure*}

\subsection{Adiabatic regime}
If $\gamma_{1,2} \simeq \gamma \ll \omega \ll \Delta E$, the quantum states evolve adiabatically following the changing flux. However, since population relaxation occurs at a timescale $\gamma^{-1}$, the state populations do not respond to the oscillating flux and the inverse inductance matrix in this limit is expressed as 
\begin{equation}\label{eq:adiabatic}
    K_{\alpha \beta} = \sum_{i} \rho_{i}\partial_{\Phi_\beta} \langle  \hat{I}_\alpha \rangle_{i} = \sum_{i} \rho_{i}\partial_{\Phi_\alpha}\partial_{\Phi_\beta} E_i.
\end{equation}

We note that in this limit the inverse inductance matrix is symmetric even for driven system, in distinction from the zero-frequency limit.

The adiabatic inverse inductance matrix is given by the energy curvature tensor so that it should exhibit paired extrema features at avoided crossings similar to zero-frequency limit results. This is confirmed
by inspecting the Fig. \ref{fig:inductance_frequency_independent} where the blue curves show the inverse inductance matrix for the same settings that we use for the zero-frequency limit. We see that qualitatively both limits are the same, although the peak values may be suppressed by a factor in the adiabatic limit. The most notable difference is observed in the left pane of Fig. \ref{fig:inductance_frequency_independent} (b):
the anti-symmetric part of the matrix is zero.

We illustrate the detailed frequency dependence at $\omega \simeq \gamma$ in Fig. \ref{fig:inductance_frequency_dependent} plotting both real and imaginary part of $\tilde{K}$. The settings correspond to those of Fig. \ref{fig:inductance_frequency_independent} (d) while the flux is set to $\varphi = 2.4 f$ where all elements of $\tilde{K}$ are close to the maximum. The real part changes from zero-frequency limit at $\omega \to 0$ to adiabatic regime at $\omega \gg \gamma$.
The imaginary part $\propto \omega$ at $\omega \to 0$. It is expected to decrease as $\omega^{-1}$ also for  $\omega \gg \gamma$. We do not see this decrease completed in the plots since $ \gamma = 10^{-2} f$ and at largest frequencies in the plots $\omega \simeq 0.1$  we transit to another regime, the resonant regime discussed in the next subsection.

\subsection{Resonant regime}
At higher frequencies $\omega{\sim} \Delta E \gg \gamma$, the current response can be directly evaluated with the Kubo formula applied to $H_{\rm eff}$. The normalized inverse inductance matrix then reads:
\begin{equation}\label{eq:ressonant}
    \tilde{K}_{\alpha \beta}(\omega) = \sum_{i j}(\rho_{i} - \rho_{j}) \frac{\bra{j}{\sigma}^z_\alpha \ket{i} \bra{i} {\sigma}^z_\beta \ket{j}}{\omega - E_i + E_j+ i0},
\end{equation}
where the eigenstates $\ket{i}$, their eigenvalues $E_i$, and state populations $\rho_i$ are those of the unperturbed system. 
For $\omega {\rightarrow} 0$ this reduces to Eq.~\eqref{eq:adiabatic} that describes the adiabatic regime. However, the resulting expression does not agree with one at zero-frequency limit where the response of the state populations is important.

As function of frequency, the imaginary part of the symmetric part of $\tilde{K}$ is a set of delta-functions at the positions of the resonances at all possible eigenvalue differences $E_{i}-E_{j}$, while imaginary part diverges $\propto (\omega -E_{i}+E_{j})^{-1}$  at these positions. The divergences at the resonant positions are smoothed at the frequencies $|\omega {-} (E_i {-} E_j)|{\sim} \gamma$ where Eq.~\eqref{eq:ressonant} is not applicable. 
so that the delta-functions become Lorentzian peaks and real part exhibits paired extrema features. The scale of $\tilde{K}$ is $\gamma^{-1}$ in the vicinity of the resonances and $(\Delta E)^{-1}$ otherwise. As to anti-symmetric part of $\tilde{K}$, it is just opposite: real part gives peaks while imaginary part gives paired extrema features. 

To describe the response beyond Eq.~\eqref{eq:ressonant}, and reproduce the smoothing of the resonances, one solves the time-dependent Lindblad equation in the limit of small oscillating fluxes. This is how we produce the illustrative results presented in Fig.~\ref{fig:inductance_resonant}. We choose the settings to achieve good hybridization so that diagonal and non-diagonal elements of $\tilde{K} f$ are of the same order of magnitude. Indeed we observe the resonances in imaginary part and paired extrema features in real part of the symmetric part (Fig.~\ref{fig:inductance_resonant} (a),(b),(c)), and opposite features in the anti-symmetric part (Fig.~\ref{fig:inductance_resonant} (d)) at all eigenvalue differences. The anti-symmetric part is non-zero because the system is driven.

\section{\label{sec:two_tone}Resolving AM states with two-tone spectroscopy}

Within the circuit QED paradigm, the standard way to read out qubit states involves coupling of a qubit to an oscillator with a frequency $\omega_{\rm osc}$ which is different from the qubit resonant frequency $2E_A$~\cite{blaisCircuitQuantumElectrodynamics2021}. While the qubit is driven at frequency $\Omega_1 \approx 2E_A$, the oscillator is driven at different frequency $\Omega_2 \approx \omega_{\rm osc}$. This is why it can be called two-tone spectroscopy.  As discussed in Sec. \ref{sec:Setup}, it results in a state-dependent frequency shift of the oscillator.
In this Section, we illustrate the use of two-tone spectroscopy for measuring the AM states in our excited two-ALQ setup.

In Appendix \ref{sec:appendixB} we provide an extensive derivation of Lindblad equation for the situation of two-tone spectroscopy. We did it for arbitrary numbers of qubits and oscillators, and take into account dissipative parts in the response functions. 
It involves a RWA with a frame rotating at $\Omega_1$ for the qubit operators and another frame rotating at $\Omega_2$ for the creation and annihilation operators of the oscillators. Thereby we neglect terms oscillating at the large frequencies $\sim \Omega_{1, 2}$, but retain the coupling between the qubits and oscillators oscillating at the beating frequency $|\Omega_1 - \Omega_2|$. The dissipation eventually complicates the Lindblad equation giving rise to non-trivial non-Hamiltonian terms that involve the qubit operators along with the boson creation/annihilation operators. While this could lead to interesting effects, the simplest and most practical limit is that of purely imaginary impedance.

In this limit, the most important extra terms are of Hamiltonian form. We take two ALQs and one oscillator to arrive at 
\begin{equation}
    H_{\rm tt} = H_{\rm eff} + 
	(\Delta - \kappa_1 \sigma_1^z -\kappa_2 \sigma_2^z)b^\dagger b 
	+ A_{\text{osc}} (b + b^\dagger).
\end{equation}
Here, $\Delta = \omega_{\rm osc}-\Omega_2$ is the detuning of the second drive from the oscillator frequency, $A_{\text{osc}}$ is the oscillator drive amplitude in frequency units. Alternatively we can write
\begin{align}
    H_{\rm tt} &= \sum_{n} (H^{(n)}_{\rm tt} + \Delta n )|n\rangle\langle n| + A_{\text{osc}} (b + b^\dagger),  \\
	H^{(n)}_{\rm tt} &\equiv H_{\rm eff} -n(\kappa_1 \sigma_1^z+ \kappa_2 \sigma_2^z),\label{eq:AC}
\end{align}
to make explicit that a certain number of photons $n$ in the oscillator effectively shifts the ALQ frequencies~\cite{autlerStarkEffectRapidly1955}.

We assume the oscillator is connected at port "3" in our setup. The state dependent frequency shift of the oscillator is then given by Eq.~\eqref{eq:oscillator_state_dependent_shift} of the circuit theory analysis
\begin{equation}
    \kappa_{\alpha} = \bar{Y}_{\alpha}\bar{Y}_{\text{osc}}\Im \{Z_{3 \alpha}\} \Im \{Z_{\alpha 3}\} \frac{\Omega_1  \Omega_2}{\Omega_2 - \Omega_1}.
\end{equation}
This expression is valid for $|\Omega_2 -\Omega_1| \ll \Omega_{1,2}$ and negligible frequency dependence of the impedance between the oscillator and ALQs.
The Lindblad equation is obtained by adding the ALQ decay terms with $\gamma_{1,2}$ and the dissipative term  $2\gamma_{\text{osc}} \mathcal{D}(b, b^\dagger)$ to account for finite line-width of the oscillator. 
Note that $\lambda$ in $H_{\rm eff}$ is also enhanced by the oscillator,
\begin{align}
\lambda = \lambda_0 + \tilde{\lambda};\; \tilde{\lambda} = \sqrt{\bar{Y}_{1} \bar{Y}_{2}} \Im \{Z_{3 1}\} \bar{Y}_{\text{osc}} \Im \{Z_{2 3}\} \frac{\Omega_1  \Omega_2}{\Omega_2 - \Omega_1}
\end{align}
As noted in Sec. \ref{sec:Setup}, generally $\kappa_\alpha$ and $\lambda$ are of the same scale.
Moreover, there is an exact relation between $\kappa_{1,2}$ and the enhanced part of $\lambda$ in our setup, 
\begin{equation}
\kappa_1 \kappa_2 = \tilde{\lambda}^2.
\end{equation}

This is disadvantageous if two-tone spectroscopy is used for the measurement of the ALQ system. For a non-obtrusive measurement, one has to require $\kappa_{\alpha} \ll \Delta E$, that is, the modification of $H_{\rm eff}$ by a single photon in the oscillator is small. Since in interesting cases $\Delta E \simeq \lambda$, the requirement is far from being fulfilled automatically. However, a bit of circuit design helps with this: one can increase $\lambda_0$ keeping $\tilde{\lambda}$ small (in~\cite{cheungPhotonmediatedLongrangeCoupling2024} this has been achieved by using one oscillator mode for the enhancement and another one for the two-tone measurement). At any rate, the terms with $\kappa_{\alpha}$ provide effective detuning of the ALQ system (see Eq.~\eqref{eq:AC}).
 This implies that the number of photons in the oscillator should not exceed much $\lambda/\kappa_\alpha$, and therefore a proportionally small value of the oscillator response. The experience of~\cite{cheungPhotonmediatedLongrangeCoupling2024} proves that the small value of the response is not a problem for an efficient measurement.

Let us concentrate on linear response $\langle b \rangle \propto A_{\rm osc}$. We identify three interesting regimes. In a single-peak regime $\Delta E\gg \kappa$, $\Delta E \gg \gamma_{1,2} \gg \gamma_{\rm osc}$,  the dynamics of ALQ system are faster than equilibration of the oscillator response. Therefore, the individual frequency shifts of four dressed states $z_i {\equiv} \bra{i}\kappa_1 \sigma_1^z {+}\kappa_2 \sigma_2^z \ket{i}$ are averaged with the weights of corresponding populations $\rho_i$, 
and the response is given by 
\begin{equation}\label{eq:oscillator_current_1}
    \langle b \rangle = A_{\text{osc}} \frac{1}{-\Delta + \bar{z}  +i \gamma_{\text{osc}}},
\end{equation}
where the shift of the resonant peak is $\bar{z} = \sum_{i} \rho_i z_i$. In principle, this is enough for measurements and characterizations.

In a 4-peak regime $\Delta E \gg \kappa \gg \gamma_{\rm osc} \gg \gamma_{1,2}$, the dynamics of of ALQ system are slower than the equilibration of the oscillator, and it is possible to resolve the individual frequency shifts of all four states from the positions of four distinct peaks,
\begin{equation}\label{eq:oscillator_current}
    \langle b \rangle = A_{\text{osc}}\sum_j \frac{\rho_j}{-\Delta + z_j   +i \gamma_{\text{osc}}}
\end{equation}
The heights of the peaks are given by the corresponding populations, and $\bar{z}$ can be recovered by integration of imaginary part of the response over $\Delta$.
This regime provides most direct information about the states of ALQ system and their populations. 

In a 4$\times$4-peak regime, $\Delta E \simeq \kappa \gg \gamma_{\rm osc} \gg \gamma_{1,2}$, the peaks in linear response indicate the transitions between any of the four states of $H^{(0)}_{\rm tt}$ (no photons) to any of the four states of $H^{(1)}_{\rm tt}$ (a single photon), 
\begin{equation}\label{eq:oscillator_current-2}
    \langle b \rangle = A_{\text{osc}}\sum_{j,j'} \frac{\rho_j |\langle j|j'\rangle|^2}{-\Delta + E'_{j'} - E_j   +i \gamma_{\text{osc}}}
\end{equation}
where $j$ labels the states of $H^{(0)}_{\rm tt}$ with eigenvalues $E_j$ while $j'$ labels the states of $H^{(1)}_{\rm tt}$ with eigenvalues $E'_{j'}$. In the limit $\kappa \ll \Delta E$, $|\langle j|j'\rangle|^2 \to \delta_{jj'}$, $E'_j - E_j = \bra{j}\kappa_1 \sigma_1^z {+}\kappa_2 \sigma_2^z \ket{j}$: this reproduces Eq.~\eqref{eq:oscillator_current}.  

We provide numerical illustrations for 4-peak and 4$\times$4-peak regimes, also beyond the linear response.

\begin{figure}
    \includegraphics[width=\linewidth]{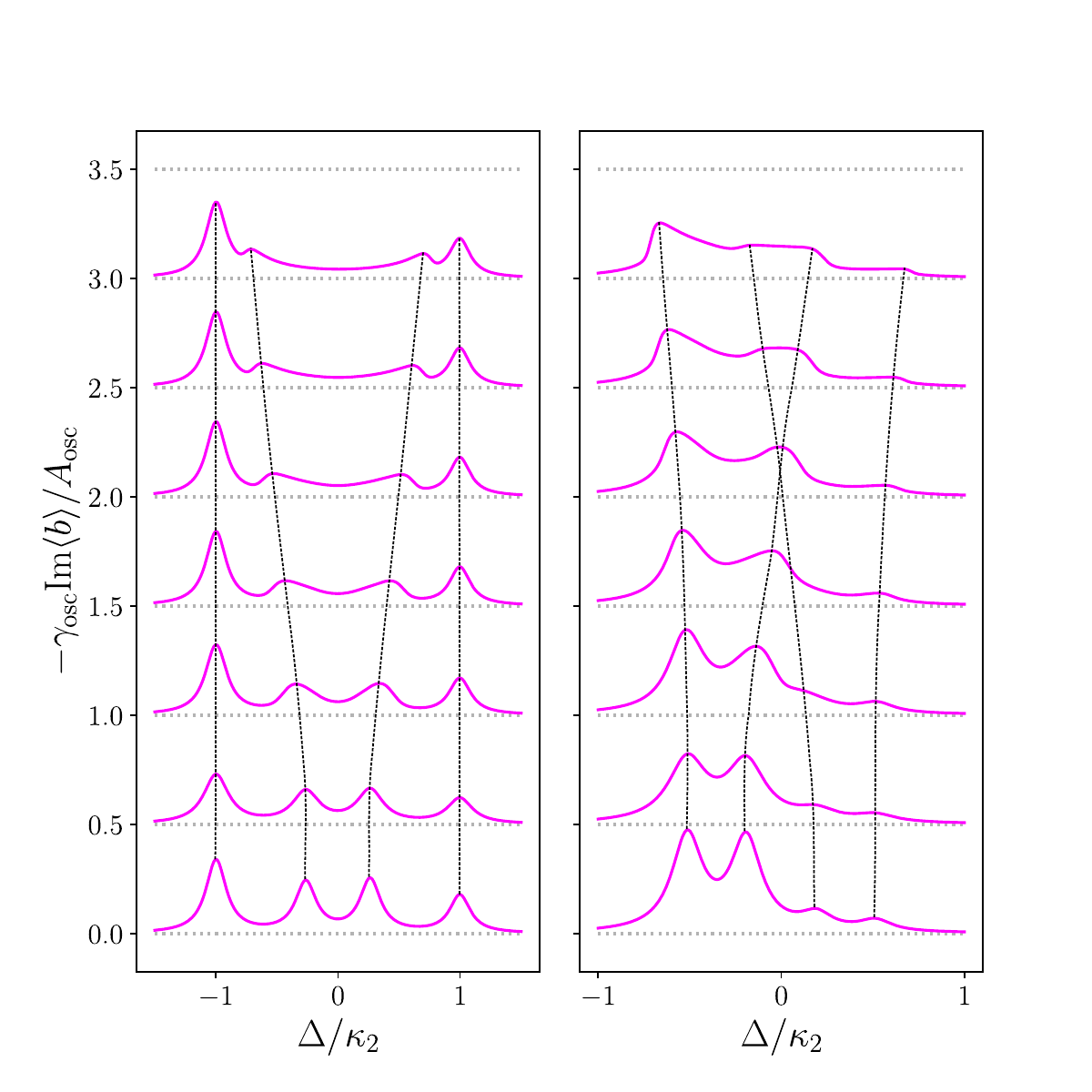}
    \caption{\label{fig:two_tone} 
	Two-tone spectroscopy in the 4-peak regime. The normalized dimensionless oscillator response $-\gamma_{\rm osc}\Im \langle b\rangle /A_{\text{osc}}$ is plotted versus the detuning $\Delta$ at a set of drive values $A_{\rm osc}/\gamma_{\rm osc} = [0.01, 0.5, 1, 1.5, 2, 2.5, 3]$ for two sets of parameters. The plots are offset by $1/2$ for clarity. The curved dashed lines are guides for the eye indicating the shifts of peak positions upon changing  $A_{\rm osc}$. For both sets, $\kappa_2=0.1 \lambda, \kappa_1=0$, $\gamma_{\rm osc} = 0.1 \kappa_2$, and $\theta=0$. Left pane: flux-tuned resonance (c.f. Fig. \ref{fig:small_interction_steadystate}). We choose $\varphi =-f$, $A_2=0$, $A_1 = 2.02 f$ so we are slightly off the resonance, $\varphi_c + \varphi = 0.02 f$, $\lambda= 0.2 f$.
	The ALQ decay rates are: $\gamma_2 =\gamma_{\rm osc}/500$, $\gamma_1 =\gamma_{\rm osc}/50$.
	Two peaks in the middle correspond to the AM states formed close to the resonance. Upon increasing $A_{\rm osc}$, their hybridization and population ceases owing to AC Stark detuning.
	Right pane: degenerate qubits, $\varphi=0$. The parameters are: $f=0.3 \lambda$, $A_1=A_2=0.7 \lambda$, $\gamma_{1,2} =\gamma_{\rm osc}/50$. We choose sufficiently strong driving to excite all 4 states. The shift of peak positions with increasing $A_{\rm osc}$ indicates the modification of the eigenstates, the change of their heights indicates the change of the eigenstate populations. 
	}\end{figure}

In  Fig. \ref{fig:two_tone} we illustrate the 4-peak regime. 
We plot the imaginary part of the normalized oscillator response, $-\Im \langle b\rangle / A_{\text{osc}}$ at various values of $A_{\text{osc}}$, since the peaks in this quantity indicate resonant positions as function of the detuning $\Delta$. The widths of the peaks are $\simeq \gamma_{\rm osc}$. Here we take $\kappa_2=0.1 \lambda$ and  $\kappa_1=0$, so that the enhancement $\tilde{\lambda}=0$, and the oscillator is only coupled to the first ALQ. In this case, the presence of four distinct peaks immediately signifies hybridization. Indeed, if the four eigenstates are product states of two ALQs, they give rise to pairwise coinciding peak positions. Only two peaks would thus be visible. The positions of the peaks in $\Delta$ immediately give $\langle \sigma^z_2 \rangle$ in the corresponding states.

In the left pane, we address the flux-tuned resonance (c.f. Fig. \ref{fig:small_interction_steadystate}) driving the first ALQ only ($A_2=0$). The parameters are slightly off the resonance, $(\varphi_c - \varphi)/\varphi = 0.02$. We also choose $\gamma_1 = 10 \gamma_2$ to assure comparable population of both states of the second ALQ despite the fact that its state is hardly dressed. In this case, we expect the states
$|e'g\rangle$, $|g'e\rangle$ to hybridize at the resonance while the states $|e'e\rangle$, $|g'g\rangle$ are not hybridized but also populated. In the Figure, the hybridized states give two peaks in the middle. They would give the same $\langle \sigma^z_2\rangle =0$ exactly at the resonance, so we tune the parameters slightly off the resonance to have them split. The peaks at $\Delta =\pm \kappa_2$ correspond to the product states $|e'e\rangle$, $|g'g\rangle$ with $\langle \sigma^z_2\rangle =\pm 1$. We observe that increasing the drive detunes the ALQ system from the resonance. This breaks up the hybridization of $|e'g\rangle$, $|g'e\rangle$ so the positions of the corresponding peaks move towards $\pm \kappa_2$. Their population decreases as seen from the peak heights.

In the right pane, we illustrate the 4-peak regime for degenerate qubits. We choose $f=0.3$ between the positions of hybridization and two-photon resonance and choose sufficiently big drive to achieve noticeable population of all 4 states. At small drive, the peak positions correspond to expected $\langle \sigma^z_2 \rangle$ in all four states. All states are noticeably hybridized. The two least populated states (two peaks on the right) have large fraction of $|ee\rangle$ as expected. Correspondingly, the leftmost peak has largest fraction of $|gg\rangle$. The positions of the peaks are shifted upon increasing the drive indicating the modification of the eigenfunctions by AC Stark effect. In a very simplistic picture, the wavefunctions can be obtained from the Hamiltonian of Eq.~\eqref{eq:AC}  by replacing $n$ with effective $\bar{n} \simeq (A_{\rm osc}/\gamma_{\rm osc})^2$ for each value of the drive. This picture does not quite work. Firstly, the average number of photons does depend on $\Delta$ for each $A_{\rm osc}$: it is proportional to the square of the response and is thus higher at the peaks. Secondly, the average number of photons is not a relevant number for the modification. At the position of the peak,  the ALQ system switches between the states. The number of photons is high for the state corresponding to the peak and close to 0 for all other states, so the averaged number of photons is smaller than the relevant one. 
However, this simple picture works qualitatively. We see the leftmost peak approaching $\langle \sigma^z_2 \rangle = -1$ upon increasing the drive. This indicates that the state is close to $|gg\rangle$. Its height increases because the detuning of the ALQ system which makes $|gg\rangle$ an eigenstate and the most populated one. The rightmost peak approaches $\langle \sigma^z_2 \rangle = 1$ indicating $|ee\rangle$. The middle peaks cross upon increasing $A_{\rm osc}$: we expect this crossing at $\bar{n}\approx 2.5$ in the simple picture. For the upper plot corresponding to $A_{\rm osc}=3$, we estimate $\bar{n} \approx 8$ for the leftmost and rightmost peak, and $\bar{n} \approx 5$ for middle peaks, which is larger than the average number of photons at the peak positions. 

\begin{figure}
    \includegraphics[width=\linewidth]{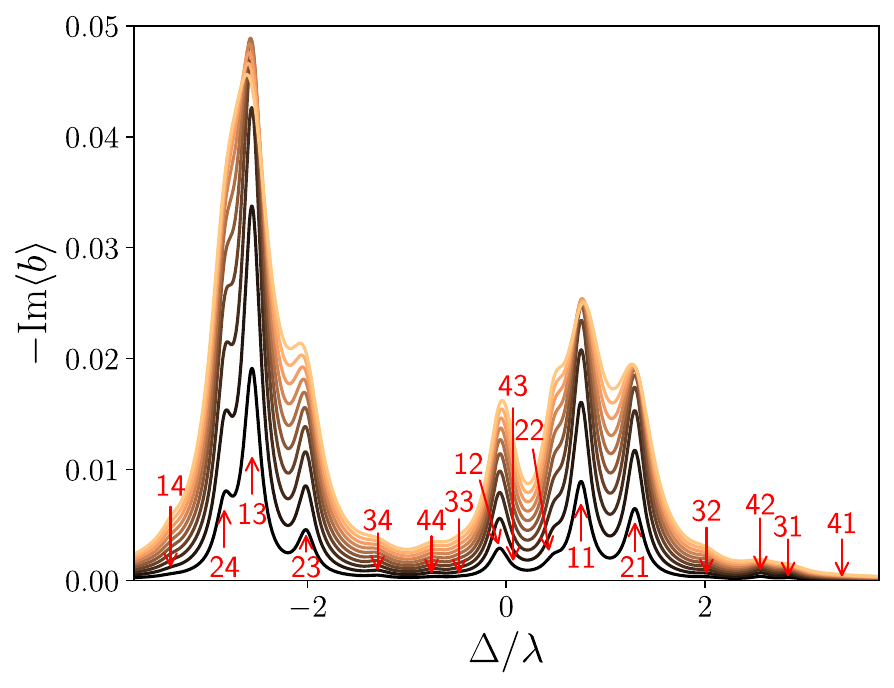}
    \caption{\label{fig:four-by-four} 
    Two-tone spectroscopy in the 4$\times$4-peak regime. The non-normalized response $-\Im \langle b\rangle$ is plotted versus the detuning $\Delta$ at a set of drive values $A_{\rm osc}$ ranging from $0.05\gamma_{\rm osc}$ to $0.5 \gamma_{\rm osc}$ in steps of $0.05\gamma_{\rm osc}$. Brighter lines correspond to larger values of $A_{\rm osc}$. The resonant peaks are due to transitions between eigenstate $\ket{j}$ of $H_{tt}^{(0)}$ and eigenstate $\ket{j^\prime}$ of $H_{tt}^{(1)}$ (see Eq.~\eqref{eq:oscillator_current-2}). In red we label the peaks "$j j^\prime$" to show the corresponding transition, where the eigenstates are ordered by increasing eigenvalues $E_j, E_{j^\prime}^\prime$. We choose $\kappa_1=0.5\lambda$, $\kappa_2=2\lambda$, $\gamma_{\rm osc}=0.1\lambda$, and $\gamma_1=\gamma_2=\gamma_{\rm osc}/50$ to ensure that we are in the 4$\times$4-peak regime. For the ALQ system we choose the remaining parameters corresponding to degenerate qubits at the hybridization point. The parameters are: $\varphi=0$, $f=0.5 \lambda$, $A_1=0.4\lambda$, and $\theta=0$. Because of large variation in peak height and overlapping tails many of the peaks are only barely distinguishable, and  the peaks "$43$" and "$41$" are not resolved in the plot.
}\end{figure}

In Fig. \ref{fig:four-by-four}, we illustrate the $4\times4$ regime.
We take here $\lambda$ to be dominated by the oscillator-induced enhancement, $\lambda \approx \tilde{\lambda}$, so that $\kappa_2\kappa_1 = \lambda^2$ and the ALQ Hamiltonian is essentially modified even by the presence of a single photon. We choose $\kappa_1 = 0.5 \lambda, \kappa_2 = 2 \lambda$, and the parameters corresponding to degenerate qubits: $\varphi=0$, position $f$ at the hybridization resonance $f =0.5 \lambda$, and drive only the first qubit with $A_1=0.4\lambda$.  
We plot the non-normalized response $- {\rm Im} \langle b\rangle$
at various drive amplitudes. At small amplitudes, the response is described by Eq.~\eqref{eq:oscillator_current-2}: we thus expect 16 peaks at the positions given by the differences of eigenvalues with zero and one photon. We see 14 in the plot. The peaks substantially vary in height due to varying overlaps between the states and different populations, so the remaining 2 are too small to see at the plot scale. In distinction from 4-peak regime, the positions of the peaks do not change with increasing $A_{\rm osc}$ as expected. We note the saturation of the peaks at $A_{\rm osc} \simeq \gamma_{\rm osc}$. Indeed, in this regime the resonance occurs between only a pair of states $n=0,1$ while the transitions to higher $n$ are not at the resonance. So the response is saturated as a response of a qubit upon increasing the drive. Note that the average number of photons as estimated from the $\langle b \rangle$ plotted is much smaller than $\langle n \rangle = 1/2$ expected at the saturation. This is due to the phenomenon mentioned in the discussion of 4-peak: the system switches between the states so number of photons is high in the resonant state and almost zero in all other states.

\section{\label{sec:Conclusion}Conclusions}

To conclude, in this paper we thoroughly discuss the opportunities to create Andreev molecules at long distance between the junction. This is in contrast to the setups where the Andreev molecules are formed by at short distances by coherent electron transfer between the junctions, the latter requires challenging nanofabrication. 

We demonstrate that the excited quasi-particle singlets in two Josephson junctions can be hybridized by photon exchange via an embedding electric circuit. The hybridization is the strongest when the energies of ABS in two junctions are aligned, this results in level repulsion. We show that the resulting energy splitting can be evaluated from the simple circuit theory and give the estimations of the effect (Sec.~\ref{sec:Setup}) . The simplest estimation is that the relative splitting is of the order of the transresistance between the junctions in units of resistance quantum. The splitting is enhanced if the resonant frequency of the junctions is close to that of an oscillator mode in the circuit. Similar configuration has been realized in ~\cite{cheungPhotonmediatedLongrangeCoupling2024}.

To get a stationary response witnessing the molecular states, one has to keep the system excited. The resonant excitation of the ABS is a straightforward and convenient way to do this. To this end, we derive the associated Lindblad equation corresponding to the persistent resonant excitation (Sec.~\ref{sec:Microscopic_model}). This gives us the tunable 4-state resonant spectrum of the molecule as well as the dissipative dynamics of the transitions between the states. We identify interesting regimes of the spectrum corresponding to small-splitting level crossings of these states.  We compute stationary populations of the ALQ states to demonstrate resonant features that reveal the hybridized superpositions. (Sec.\ref{sec:Steady_state})

We regard the low-frequency mutual inductance of the Josephson junctions as an important manifestation of non-local Josephson effect and compute all four components of the inductance matrix versus flux and/or drive frequency in various frequency regimes: zero-frequency regime, adiabatic regime, resonant regime. We find the peak structures with the enhanced value of the inverse inductance in all regimes, this should make the effect easily observable. 

Linear oscillators are commonly employed for qubit readout and manipulation. Therefore we specifically address the case where the embedding circuit contains the oscillating modes.
We derive the Lindblad equation for the conditions of two-tone spectroscopy, where the tones match the resonant frequencies of the ALQs and the oscillators. In Appendix B we present its rather complex form for general impedance circuit, while for numerical results we resort to non-dissipative imaginary impedance.

We demonstrate that the molecular states can be resolved from the oscillator response. This is provided small values of the oscillator drive and sufficiently weak coupling between the oscillator and ALQs, which requires separate oscillators used for state readout and hybridization tuning. At larger drive amplitudes the AC Stark effect detunes the ABS preventing the formation of the hybridized states.

\begin{acknowledgments}
The authors would like to thank C. Sch\"onenberger, L. DiCarlo, V. Fatemi, and L. Glazman for helpful discussions related to the project.

Code to reproduce the figures in this article is available at Zenodo~\cite{samuelsenCodeFiguresData2025}
\end{acknowledgments}

\bibliography{references_andreev_molecules}

\appendix

\section{\label{sec:appendix_A} Derivation of the Lindblad equation for ALQs coupled through a linear electric circuit} 

In this Appendix, we derive the Lindblad equation for a set of quantum systems embedded into an electromagnetic environment. While standard, this derivation is important for the main text, in particular, it helps to understand the efficiency of circuit theory analysis for the evaluation of "quantum" effects.

We start with a Hamiltonian encompassing a set of quantum systems labeled by $\alpha$ and embedded into electromagnetic environment. We assume linear coupling of a set of flux operators $\hat{\Phi}_\alpha$ in the environmental degrees of freedom with corresponding  
\begin{equation}
H= H_\alpha - \hat{I}_\alpha (\Phi^{d}_\alpha(t) + \hat{\Phi}_\alpha) + H_{{\rm env}}
\end{equation} 
We thus consider linear coupling to a set of fluxes, $\hat{I}_\alpha$ being the corresponding current operators in the space of the system degrees of freedom. 

In addition, we assume that each system is close to resonance at frequency/energy $E$ so that, each system comprises several groups of states separated by a "big" energy difference $E$
\begin{equation}
H_0 = \bar{H} + E \Sigma,
\end{equation}
where the eigenvalues of $\Sigma$  are positive integers. For a qubit, these integers are $0$ and $1$, while for an oscillator they can be any numbers.

The systems are driven at frequency $\omega$ close to $E$, that is,
\begin{equation}
\Phi^{d}_\alpha(t) = \Phi^{d}_\alpha e^{-i\omega t} + \Phi^{d*}_\alpha e^{i\omega t}
\end{equation}
We perform a rotating wave transformation with this frequency and matrix $\Sigma$. Most generally, the current operators are transformed to 
\begin{align}
\hat{I}_\alpha \to \hat{J}_\alpha + \hat{I}^+_\alpha e^{i \omega_2 t } + \hat{I}^-_\alpha e^{-i \omega_2 t} + \nonumber \\
 \sum_{n=2} \left(\hat{I}^{n,+}_\alpha e^{i \omega n t } + \hat{I}^{n,-}_\alpha e^{-i \omega n t}\right).
\end{align}  
Here, the operators $\hat{J}_\alpha$ are diagonal within the groups of the states with different eigenvalues of $\Sigma$. The operators $I^{+,-}$ are non-diagonal ($I^+= (I^-)^\dagger$) describing the transitions between the neighboring groups and oscillate at main frequency. There are also terms oscillating at higher harmonics but they are canceled by RWA approximation so that the time-averaged RWA Hamiltonian reads
\begin{equation}
H_{\rm RWA} = \bar{H} - f \Sigma - \hat{I}^+_\alpha  \Phi^{d}_\alpha - \hat{I}^-_\alpha \Phi^{d*}_\alpha,
\end{equation}
$f\equiv \omega-E$.

Next, we compute the effect of the environmental fluctuations in the second order. We concentrate on a part of the environment at frequencies $\approx E$ and thus neglect the terms coming with $\hat{J},I^{n,+}_\alpha$. The derivation is rather standard yet we do it in some detail. 
The perturbation is thus
\begin{equation}
H'(t) = - I_\alpha^+ e^{i E t} \hat{\Phi}_\alpha(t)  - I_\alpha^- e^{-i E t} \hat{\Phi}_\alpha(t) 
\end{equation}
where we have neglected the difference between $E$ and $\omega$ assuming the correlators do not change much at this frequency scale.
The second order terms read
\begin{align}
\left(\partial_t \rho\right)_{\rm (2)} = \int_{-\infty}^t dt'   \left( - H'(t) H'(t') \rho(t')- \right. &&\left. \rho(t') H'(t') H'(t) \right. \nonumber \\
\left. H'(t) \rho(t') H'(t') + H'(t') \rho(t') H'(t) \right)&&
\end{align}
We assume that $\rho$ does not change at the time scale $1/E$ and thus neglect its time dependence under the integral, $\rho(t') \approx \rho(t)$. We introduce the correlators of flux operators
\begin{align}
S^{A,R}(t) = \Theta(\pm t) \langle \Phi_\alpha(0) \Phi_\beta(t)\rangle \\
S^{R\mp}_{\alpha\beta} \equiv S^{R}_{\alpha\beta}(\mp E) =\int\limits_{-\infty}^t dt' \langle\Phi_\alpha(t) \Phi_\beta(t') \rangle e^{\pm i E(t-t')} ;\\
S^{A\pm}_{\alpha \beta} = \equiv S^{A}_{\alpha\beta}(\pm E)\int\limits_{-\infty}^t dt' \langle\Phi_\alpha(t') \Phi_\beta(t) \rangle e^{\pm i E(t-t')} 
\end{align}

We rewrite the second-order terms in terms of these correlators
\begin{align}
\left(\partial_t \rho\right)_{\rm 2nd} = 
S^{R-}_{\alpha \beta} (-\hat{I}^+_\alpha \hat{I}^-_\beta \rho +
  \hat{I}^-_\beta \rho{\hat I}^+_\alpha) &&\nonumber\\
+ S^{R+}_{\alpha \beta} (-\hat{I}^-_\alpha \hat{I}^+_\beta \rho + \hat{I}^+_\beta \rho\hat{I}^-_\alpha)  
+S^{A+}_{\alpha \beta} (- \rho\hat{I}^-_\alpha \hat{I}^+_\beta &+
& \hat{I}^+_\beta \rho \hat{I}^-_\alpha) \nonumber\\
+ S^{A-}_{\alpha \beta} (-\rho\hat{I}^+_\alpha \hat{I}^-_\beta  + \hat{I}^-_\beta \rho \hat{I}^+_\alpha), && \label{eq:firstrun}
\end{align}
and regroup separating the terms into dissipative ones and corrections to the Hamiltonian. Let us introduce a notation
\begin{equation}
{\cal D}(\hat{A},  \hat{B}) \equiv \hat{A} \rho \hat{B} - \frac{1}{2} \left(\hat{\rho} \hat{B} \hat{A} + \hat{B} \hat{A} \hat{\rho}\right).
\end{equation}
Defining $S_{\alpha\beta} = S_{\alpha\beta}^{A} + S_{\alpha\beta}^{R}$, we rewrite Eq.~\eqref{eq:firstrun} as 
\begin{align}
\label{eq:diss}
\left(\partial_t \rho\right)_{\rm (2)} = 
S^-_{\alpha\beta} {\cal D}(\hat{I}^{-}_\beta,  \hat{I}^{+}_\alpha) 
+ S^+_{\alpha\beta} 
{\cal D}(\hat{I}^+_\beta,  \hat{I}^{-}_\alpha)
- i [H_{\rm (2)}, \rho];  \\
H_{\rm(2)}= \frac{i}{2} \left( (S^{A-}_{\alpha\beta} - S^{R-}_{\alpha\beta}) \hat{I}^+_{\alpha} \hat{I}^-_\beta + (S^{A+}_{\alpha\beta} - S^{R+}_{\alpha\beta}) \hat{I}^-_{\alpha} \hat{I}^+_\beta\right)
\end{align}

Next, we express the correlators with using the response function $\chi_{\alpha\beta}(\omega) = i Z_{\alpha\beta}(\omega)/\omega$, $\chi_{\alpha\beta}(-\omega) = \chi_{\alpha\beta}^*(\omega)$ The fastest way is to represent the flux operators as linear superpositions of bosonic creation and annihilation operators, choosing the coefficients to mimics the response function. We assume that the boson filling factor $n_B(\omega)$ depends on frequency only. If the environment is in a thermal equilibrium with the temperature $T$, $n_B(\omega) = 1/\exp(\hbar \omega/k_BT)$. It is convenient to define an odd-in-frequency function $F(\omega)$,
\begin{equation}
F(\omega) = 1+2n_B(\omega) \ {\rm at} \ \omega>0; F(-\omega)=-F(\omega). \\
\end{equation}
With this, the Fourier component of $S_{\alpha\beta}$ reads
\begin{equation}
S_{\alpha\beta}(\omega) = \frac{i}{2} (\chi_{\beta\alpha}(\omega) - \chi^*_{\alpha\beta}(\omega)) (1 -F(\omega))
\end{equation}
We also need Fourier transforms of the advanced and retarded correlators,
\begin{align}
&S^{A,R}(\omega) =  \int dt e^{i\omega t} \Theta(\pm t) S_{\alpha\beta}(t) \nonumber \\
&= \int \frac{d \nu}{2 \pi} \frac{\mp i}{\omega-\nu \pm i\delta} S_{\alpha\beta}(\nu),
\end{align}
those are analytical continuations of $S_{\alpha\beta}(\omega)$ to lower/upper plane of the complex $\omega$. It is convenient to introduce 
\begin{equation}
X_{\alpha\beta}(\omega) = \int \frac{d \nu}{2 \pi} \frac{ \chi_{\beta\alpha}(\nu) - \chi^*_{\alpha\beta}(\nu)}{\omega-\nu + i\delta} F(\nu).
\end{equation}
$X_{\alpha\beta}(\omega) = X^*_{\alpha\beta}(-\omega)$,
that provides the upper half-plane continuation of the $F$ dependent part of $S_{\alpha\beta}(\omega)$. 
With this,  
\begin{align}
S^A_{\alpha\beta}(\omega) &= \frac{1}{2}\left( i \chi_{\beta\alpha}(\omega) + X_{\alpha\beta}(\omega) \right)\\
S^R_{\alpha\beta}(\omega) &=  \frac{1}{2}\left( - i \chi^*_{\alpha\beta}(\omega) + X^*_{\beta\alpha}(\omega)  \right)
\end{align}

Further, we denote $(\chi,X)_{\alpha\beta}(E)=(\chi^*,X^*)_{\alpha\beta}(-E) \equiv (\chi,X)_{\alpha\beta}$ and express the second-order terms as 
\begin{align}
&\left(\partial_t \rho\right)_{(2)} = -i[H^{(2)},\hat{\rho}] + \gamma_{\alpha\beta}\left((1+n(E)) {\cal D}(\hat{I}^-_\beta,\hat{I}^{+}_\alpha) \right. \nonumber \\ 
&\left. + n_B(E) {\cal D}(\hat{I}^+_\alpha, \hat{I}^{-}_\beta) \right)
\end{align} 
with  $\gamma_{\alpha\beta} = -i (\chi_{\alpha\beta} - \chi^*_{\beta\alpha})$,
\begin{align}
&H_{\rm (2)} = \frac{1}{4}\left[ - (\chi_{\alpha\beta} + \chi^*_{\beta\alpha}) (\hat{I}^+_{\alpha} \hat{I}^-_{\beta} +\hat{I}^-_{\beta} \hat{I}^+_{\alpha}) \right. \nonumber \\ &\left. - i(X_{\beta\alpha}-X^*_{\alpha\beta}) [\hat{I}^+_{\alpha}, \hat{I}^-_{\beta}] \right];\label{eq:sec-terms}
\end{align}

Perhaps more instructive representation is obtained 
if we express the current operators in terms of dimensionless raising/lowering operators, $\hat{I}_\alpha^+ = I_\alpha \hat{L}^+_\alpha$, 
$\hat{I}_\alpha^- = I_\alpha \hat{L}_\alpha$. For qubits, $\hat{L}^+,\hat{L} = \sigma_+,\sigma_-$, For oscillators, they correspond to usual bosonic creation/annihilation operators. At any rate, these operators commute for different systems, $[\hat{L}^+_\alpha,\hat{L}_\beta] = [\hat{L}_\alpha,\hat{L}_\beta]=0$.
\begin{align}
\label{eq:secondorder}
&\left(\partial_t \rho\right)_{(2)} = -i[H^{(2)},\hat{\rho}] + \Gamma_{\alpha\beta}\left((1+n(E)) {\cal D}(\hat{L}_\beta,\hat{L}^{+}_\alpha) \right. \nonumber \\ 
&\left. + n_B(E) {\cal D}(\hat{L}^+_\alpha, \hat{L}_\beta) \right); \Gamma_{\alpha\beta} = -i I_\alpha I_\beta(\chi_{\alpha\beta} - \chi^*_{\beta\alpha}),
\end{align} 
\begin{align}
&H_{\rm (2)} = \frac{1}{2} \lambda_{\alpha\beta} (\hat{L}^+_{\alpha} \hat{L}_{\beta} +\hat{L}_{\beta} \hat{L}^+_{\alpha}) + r_\alpha [\hat{L}^+_{\alpha}, \hat{L}_{\beta}]; \\\label{eq:sec-terms-2-L}
&\lambda_{\alpha\beta} = - \frac{1}{2} (\chi_{\alpha\beta} + \chi^*_{\beta\alpha}) I_\alpha I_\beta; \\
&r_\alpha = -\frac{i}{4} I_\alpha I_\beta (X_{\alpha\alpha}-X^*_{\alpha\alpha})
\label{eq:sec-terms-2}
\end{align}

All parameters $\Gamma, \lambda, r$ now have dimension of rate/energy. The terms with $\Gamma_{\alpha \beta}$ come from imaginary parts of the susceptibilities and describe the relaxation of quantum systems accompanied by absorption (factor $n_B(E)$) or emission (factor $1+n_B(E)$) of a single photon with energy $E$. The peculiarity is that the Hermitian matrix $\Gamma_{\alpha\beta}$ is not diagonal so that the relaxation of different systems is not independent. In principle, such relaxation processes in the presence of drive and/or excitations in the environment may lead to formation of observable superpositions, that is, molecular states. However, this is not in the main steam of modern quantum technologies, and experimental observation is more difficult than in the case of Hamiltonian terms. This is why we do not discuss the effect of non-diagonal $\Gamma_{\alpha\beta}$ in this article.

In the Hamiltonian part, the terms with $\lambda_{\alpha\beta}$ for $\alpha \ne \beta$ describe the environment-induced interaction between the quantum systems. The environment also modifies the Hamiltonian $H_\alpha$ for each system that is described by the terms 
$\lambda_{\alpha\alpha}, r_{\alpha}$.

To make a connection with the circuit theory consideration presented in the Section \ref{sec:Setup}, let us assume that all quantum systems are oscillators. Owing to linearity of the oscillator dynamics, the classical linear circuit theory should be precisely applicable reproducing all results concerning the frequency shifts and broadening of the oscillator modes. Therefore, the expressions for $\lambda_{\alpha\beta}, \Gamma_{\alpha\beta}$ can be readily obtained from the circuit analysis.

We note that the terms with $r_\alpha$ are not relevant for the oscillators giving rise to an insignificant constant. These terms are relevant for any other quantum system. However, they only describe individual modifications of the system Hamiltonians not riving rise to interaction, and can be therefore neglected.

To make the connection even more explicit, we note that the admittance of each  quantum system (in the ground state) has a resonant pole described by Eq.~\eqref{eq:JJ_admittance} with $\bar{Y}_\alpha = - E I_\alpha^2$. Expressing $\chi$ through impedance, we reproduce

\begin{align}
\lambda_{\alpha\beta} = -i (Z_{\alpha\beta} - Z^*_{\beta\alpha})\frac{E}{2} \sqrt{\bar{Y}_\alpha \bar{Y}_\beta};  \; \\
\Gamma_{\alpha\beta} = E \sqrt{\bar{Y}_\alpha \bar{Y}_\beta} (Z_{\alpha\beta} + Z^*_{\beta\alpha}).\label{eq:gamma_matrix}
\end{align}

\section{\label{sec:appendixB} Derivation of the Lindblad equation for coupled ALQs and oscillators in two-tone spectroscopy}

In this Appendix, we provide the derivation of Lindblad equation for the situation of two-tone spectroscopy. Such situation is common in quantum information processing where a qubit is manipulated with oscillating pulses at frequency $E$ while its state is measured with an oscillator driven at another frequency $E+\Omega$. The present derivation is more extensive than a standard one where a Hamiltonian coupling between a qubit and an oscillator is derived. We consider arbitrary number of quantum systems and rather involved effects of dissipative coupling that results in non-Hamiltonian terms.

We start with a set of quantum systems (each system can be an oscillator or a qubit) of closer resonant frequencies, labeled by Latin indices. We apply to each system the fluxes oscillating with the frequency $f_1$,
$\Phi_a(t) = \Phi_a(t) e^{-i f_1 t} + \Phi^*_a(t) e^{i f_1 t}$. 
After the RWA transformation with the frequency $f_1$, the Hamiltonian for the set of quantum systems  reads
\begin{align}
H = \sum_a \left( {\omega_a} \hat{L}^\dagger_a \hat{L}_a  \right. \\ \left.- I_a\left(\hat{L}^\dagger_a(\hat{\Phi}_a + \Phi_a(t)) + \hat{L}_a(\hat{\Phi}^\dagger_a + \Phi^*_a(t))\right) \right),
\end{align}
$\hat{L}^{\dagger}_a (\hat{L}_a)$ are raising(lowering) operators of the systems,
$\omega_a$ being the resonant frequencies counted from the RWA frequency $f_1$. The applicability of RWA implies $|\omega_a| \ll f_1$.

Next step is to take into account the environment-induced coupling between the systems. We do this in the second order as described in Appendix A to come to a Bloch equation of the following form: 

\begin{align}
\partial_t \rho = -i[H,\hat{\rho}] + \gamma_{ab}\left((1+n_B) {\cal D}(\hat{L}_b, \hat{\rho}, \hat{L}^{\dagger}_a) \right. \nonumber\\ \left. + n_B {\cal D}(\hat{L}^\dagger_a, \hat{\rho}, \hat{L}_b) \right)
\end{align} 
with $\gamma_{ab} = -i I_a I_b (\chi_{ab} - \chi^*_{ba})$,
\begin{align}
H = \omega_a  \hat{L}^\dagger_a \hat{L}_a - I_a\left(\hat{L}^\dagger_a \Phi_a(t) + \hat{L}_a \Phi^*_a(t)\right) + \nonumber \\
\frac{1}{4}\left[ - I_a I_a (\chi_{ab} + \chi^*_{ba}) (\hat{L}^\dagger_{a} \hat{L}_{b} +\hat{L}_{b} \hat{L}^\dagger_{a}) \right. \nonumber \\ \left. - i I_a^2 (X_{aa}-X^*_{aa}) [\hat{L}^\dagger_{a}, \hat{L}_{a}] \right];
\end{align}
Here we assumed that $\hat{L}^\dagger_a,\hat{L}_b$ commute if $a \ne b$. 

Let us now add the second tone with frequency $f_2 = f_1 + \Omega$ , $\Omega \ll f_1$.
\begin{equation}
\Phi_a(t) = \Phi_a^{(1)} + \Phi_a^{(2)} \exp(-i \Omega t) 
\end{equation}
In principle, we could just solve the resulting time-dependent equation. However, we proceed in a different way with the goal to derive an approximate time-independent equation. 

We separate the systems into two groups: i. those with resonant frequencies close to $f_1$. We will use Greek indices to label these systems and change $\omega_a \to \epsilon_\alpha$ ii. those with resonant frequencies close to $f_2$. We retain Latin indices for the systems of the second group. Importantly, we perform an extra RWA transformation for the systems of the second group so that the resonant frequencies $\omega_a$ are counted from $f_2$, while $\hat{L}_a^\dagger$ is replaced with $\hat{L}_a^\dagger e^{i \Omega t}$. With this, the Bloch equation can be presented in the following compact way:
\begin{align}
\label{eq:td}
\partial_t \hat{\rho}(t) = S_1(\hat{\rho}) + S_2 (\hat{\rho}) + A_+(\hat{\rho}) e^{-i\Omega t} + A_-(\hat{\rho}) e^{i\Omega t} 
\end{align}

We will specify $S_{1,2}, A_{+,-}$ below. Now we approximately solve Eq.~\eqref{eq:td} with the accuracy up to the first harmonic. So we approximate the time-dependent $\hat{\rho}$ with
\begin{equation}
\hat{\rho}(t) = \hat{\rho} + e^{i\Omega t} \hat{\rho}_+ + e^{-i \Omega t} \hat{\rho}_-.
\end{equation} 
The slowly-varying part $\hat{\rho}$ satisfies 
\begin{equation}
0\approx \partial_t \hat{\rho} =S_1(\hat{\rho}) + S_2 (\hat{\rho}) + A_+(\hat{\rho}_+)  + A_-(\hat{\rho}_-) . 
\end{equation}
As to the first harmonics $\hat{\rho}_{\pm}$, they satisfy
\begin{equation}
i \Omega\hat{\rho}_+ = A_-(\hat{\rho});\; -i \Omega\hat{\rho}_- = A_+(\hat{\rho})
\end{equation}
Here, we assume $S_{1,2}, A_{+,-} \ll \Omega$ and neglect these terms  in comparison with $\Omega$.

In this approximation, $\hat{\rho}$ satisfies 
\begin{align}
\partial_t \hat{\rho}=S_1(\hat{\rho}) + S_2 (\hat{\rho}) \nonumber \\-\frac{i}{\Omega}\left( A_+(A_-(\hat{\rho}))  - A_-(A_+(\hat{\rho}))\right).
\end{align}
This is the resulting time-independent equation. It includes the terms proportional $\Omega^{-1}$ that sets the couplings between the systems of two groups. 

Let us concentrate on these terms.  
Let us now specify the corresponding superoperators introducing some compact notations
\begin{align}
\kappa_{\alpha \beta} \equiv I_\alpha I_\beta (\chi_{\alpha \beta} + \chi^*_{\beta\alpha});\; r_\alpha \equiv i I_\alpha^2 (X_{\alpha\alpha}-X^*_{\alpha\alpha})
\end{align}
\begin{align}
S_1(\hat{\rho})= -i [H_1,\hat{\rho}] + \gamma_{\alpha\beta}\left((1+n_B) {\cal D}(\hat{L}_\beta, \hat{\rho}, \hat{L}^{\dagger}_\alpha) \right. \nonumber \\ 
\left. + n_B {\cal D}(\hat{L}^\dagger_\alpha, \hat{\rho}, \hat{L}_\beta) \right); \\
H_1 = \epsilon_\alpha  \hat{L}^\dagger_\alpha \hat{L}_\alpha - I_a\left(\hat{L}^\dagger_\alpha \Phi_\alpha^{(1)} + \hat{L}_\alpha \Phi^{(1)*}_\alpha(t)\right) \\
-\frac{1}{4}\left( \kappa_{\alpha\beta} (\hat{L}^\dagger_{\alpha} \hat{L}_{\beta} +\hat{L}_{\beta} \hat{L}^\dagger_{\alpha}) +r_a [\hat{L}^\dagger_{\alpha}, \hat{L}_{\alpha}] \right); \nonumber
\end{align}

\begin{align}
S_2(\hat{\rho})= -i[H_2,\hat{\rho}] + \gamma_{ab}\left((1+n_B) {\cal D}(\hat{L}_b, \hat{\rho}, \hat{L}^{\dagger}_a) \right. \nonumber \\ 
\left.  + n_B {\cal D}(\hat{L}^\dagger_a, \hat{\rho}, \hat{L}_b) \right); \\
H_2 = \omega_a  \hat{L}^\dagger_a \hat{L}_a - I_a\left(\hat{L}^\dagger_a \Phi^{(2)}_a + \hat{L}_a \Phi^{(2)*}_a\right)  \nonumber \\
-\frac{1}{4}\left( \kappa_{ab}(\hat{L}^\dagger_{a} \hat{L}_{b} +\hat{L}_{b} \hat{L}^\dagger_{a}) + r_a [\hat{L}^\dagger_{a}, \hat{L}_{a}] \right);
\end{align}
\begin{align}
A_+(\hat{\rho}) = -i[H_+,\hat{\rho}] + \gamma_{\alpha b}\left((1+n_B) {\cal D}(\hat{L}_b, \hat{\rho}, \hat{L}^{\dagger}_\alpha) + \right. \nonumber \\ 
\left.  n_B {\cal D}(\hat{L}^\dagger_\alpha, \hat{\rho}, \hat{L}_b) \right);\\
H_+ = - I_b\Phi^{(1)*}_b \hat{L}_b - I_\alpha\Phi^{(2)}_\alpha \hat{L}^\dagger_\alpha -\frac{1}{2} \kappa_{\alpha b} \hat{L}^\dagger_{\alpha} \hat{L}_{b}
\end{align}
\begin{align}
A_-(\hat{\rho}) = -i[H_-,\hat{\rho}] + \gamma_{a \beta}\left((1+n_B) {\cal D}(\hat{L}_\beta, \hat{\rho}, \hat{L}^{\dagger}_a) \right. \nonumber \\ 
\left.  + n_B {\cal D}(\hat{L}^\dagger_a, \hat{\rho}, \hat{L}_\beta) \right);\\
H_- = - I_b \Phi^{(1)}_b \hat{L}^\dagger_b - I_a \Phi^{(2)*}_\alpha \hat{L}_\alpha -\frac{1}{2} \kappa_{a \beta} \hat{L}^\dagger_{a} \hat{L}_{\beta}
\end{align}

It is convenient to use a pseudo-Hamiltonian representation of the Liouvillian in the Bloch equation where the pseudo-Hamiltonian works on two subspaces $1,2$; and for any two operators $\hat{U},\hat{W}$
\begin{align}
\hat{U} \rho \to \hat{A}_1 \to \hat{U} \otimes 1;\; \rho \hat{W} \to 1\otimes  \hat{W}^T; \; \nonumber \\ \hat{U} \hat{\rho} \hat{W} \to \hat{U}_1 \hat{W}_2^T \to \hat{U} \otimes \hat{W}^T
\end{align}
It is always possible to choose the raising/lowering operators to be real, so that $L^T = L^\dagger$.
With this, we rewrite $A_{\pm}$ as
\begin{align}
A_+ = i(\Phi^{(1)*}_b I_b\hat{L}_b +I_\alpha\Phi^{(2)}_\alpha \hat{L}^\dagger_\alpha \nonumber \\ +\frac{1}{2} (\kappa_{\alpha b} +i \gamma_{\alpha b} (1+2n_B) \hat{L}^\dagger_{\alpha} \hat{L}_{b})  \otimes 1  \nonumber \\
-i\  1\otimes (\Phi^{(1)*}_b I_b\hat{L}^\dagger_b +\Phi^{(2)}_\alpha I_\alpha \hat{L}_\alpha  \nonumber \\+\frac{1}{2} (\kappa_{\alpha b} -i \gamma_{\alpha b}(1+2n_B)) \hat{L}_{\alpha} \hat{L}^\dagger_{b})+ 
\nonumber\\
\gamma_{\alpha b} \left( (1+n_B)  \hat{L}_{b} \otimes \hat{L}_{\alpha} + n_B  \hat{L}^\dagger_{\alpha} \otimes \hat{L}^\dagger_{b}\right) 
\end{align} 
\begin{align}
A_- = i (\Phi^{(1)}_a I_a\hat{L}^\dagger_a +I_\beta\Phi^{(2)*}_\beta \hat{L}_\beta  \nonumber \\+ \frac{1}{2} (\kappa_{a \beta}  + i\gamma_{a \beta} (1+2 n_B))\hat{L}^\dagger_{a}\hat{L}_{\beta}) \otimes 1 \nonumber \\
-i 1 \otimes (I_a\Phi^{(1)}_a \hat{L}_a +I_\beta\Phi^{(2)*}_\beta \hat{L}^\dagger_\beta  \nonumber \\
+\frac{1}{2} (\kappa_{a \beta}  - i\gamma_{a \beta} (1+2 n_B)) \hat{L}_{a}\hat{L}^\dagger_{\beta})) +
  \nonumber \\
\gamma_{a \beta} \left( (1+n_B)  \hat{L}_{\beta} \otimes \hat{L}_{a} + n_B  \hat{L}^\dagger_{a} \otimes \hat{L}^\dagger_{\beta}\right) 
\end{align} 

Next goal is to compute $A_+ A_- - A_- A_+$. We do this in three steps by. Firstly, we give the terms with the products of non-trivial direct products,
\begin{align}
(A_+ A_- - A_- A_+)_1 = \frac{n_B(1+n_B)}{2} \left( \nonumber \right.\\ \left. 
\gamma_{a \alpha} \gamma_{\alpha b} \left( [\hat{L}^\dagger_\alpha, \hat{L}_\alpha] \otimes \{\hat{L}_b^\dagger, \hat{L}_a \}  - \{\hat{L}_b, \hat{L}_a^\dagger \} \otimes [\hat{L}^\dagger_\alpha, \hat{L}_\alpha]\right) \right. \nonumber\\
\left. - \gamma_{\alpha a} \gamma_{a \beta} \left( [\hat{L}^\dagger_a, \hat{L}_a] \otimes \{\hat{L}_\beta^\dagger, \hat{L}_\alpha \} - \{\hat{L}_\beta, \hat{L}_\alpha^\dagger \} \otimes [\hat{L}^\dagger_a, \hat{L}_a]\right) \right) 
\end{align}

Second, the products of two trivial products: if 
\begin{equation}
A_{\pm} = C_1^{(\pm)} \otimes 1 + 1\otimes (C_2^{(\pm)})^T,
\end{equation}
\begin{equation}
(A_+ A_- - A_- A_+)_2 = [C_1^{+},C_1^{-}] \otimes 1 + 1 \otimes ([C_2^{-},C_2^{+}])^T
\end{equation} 
To shorten the notations, let us introduce $\tilde{\kappa}_{ab} \equiv \frac{1}{2}\left( \kappa_{ab} +i \gamma_{ab} (1+ n_B)\right)$. With this, 
\begin{align}
[C_1^{+},C_1^{-}] =  I_a^2|\Phi^{(1)}_a|^2 [\hat{L}^\dagger_a,\hat{L}_a]   -I_\alpha^2|\Phi^{(2)}_\alpha|^2 [\hat{L}^\dagger_\alpha,\hat{L}_\alpha]  \nonumber \\
+ \frac{1}{2} \left(\tilde{\kappa}_{a\alpha}\tilde{\kappa}_{\alpha b} [\hat{L}^\dagger_\alpha,\hat{L}_\alpha]  \{\hat{L}^\dagger_a,\hat{L}_b\} 
- \tilde{\kappa}_{\alpha a}\tilde{\kappa}_{a\beta} [\hat{L}^\dagger_a,\hat{L}_a]\{\hat{L}^\dagger_\alpha,\hat{L}_\beta \} \right) 
\nonumber \\
I_a [\hat{L}^\dagger_a,\hat{L}_a] \left( \Phi_a^{(1)*} \tilde{\kappa}_{a\beta} \hat{L}_\beta + \Phi^{(1)}_a \tilde{\kappa}_{\beta a} \hat{L}^\dagger_\beta\right) \nonumber \\
- I_\alpha [\hat{L}^\dagger_\alpha,\hat{L}_\alpha] \left( \Phi_\alpha^{(2)*} \tilde{\kappa}_{\alpha b} \hat{L}_b + \Phi^{(2)}_\alpha \tilde{\kappa}_{b\alpha} \hat{L}^\dagger_\beta\right)
\end{align}

and $[C_2^{+},C_2^{-}]$ is obtained from the above expression by replacing 
$\tilde{\kappa}_{ab} \equiv \frac{1}{2}\left( \kappa_{ab} +i \gamma_{ab} (1+ n_B)\right) \to \frac{1}{2}\left( \kappa_{ab} - i \gamma_{ab} (1+ n_B)\right)$.

At the third step, we collect the cross-products of trivial and non-trivial products,
\begin{align}
(A_+ A_- - A_- A_+)_3 = \\
 i \gamma_{\alpha b} \left( - \Phi^{(1)}_b I_b \left( \bar{n}_B [b] \otimes \hat{L}_\alpha + n_B \hat{L}^\dagger_\alpha \otimes [b] \right) + \right. \nonumber \\ \left.\Phi^{(2)*}_\alpha I_\alpha \left(\bar{n}_B \hat{L}_b \otimes [\alpha]  + n_B [\alpha] \otimes \hat{L}^\dagger_b\right) + 
 \right. \nonumber \\ \left. 
 - \bar{n}_B \tilde{\kappa}_{b\beta} [b] \hat{L}_\beta \otimes \hat{L}_\alpha +
 n_B \tilde{\kappa}_{a\alpha} [\alpha] \hat{L}^\dagger_a \otimes \hat{L}^\dagger_\beta + \right. \nonumber \\ \left.
 \bar{n}_B \tilde{\kappa}^*_{ a\alpha}  \hat{L}_b \otimes \hat{L}_a [\alpha] -  
  n_B \tilde{\kappa}^*_{b\beta}  \hat{L}^\dagger_\alpha \otimes \hat{L}^\dagger_\beta [b] \right) 
 \nonumber \\ 
  i \gamma_{b\alpha} \left( - \Phi^{(1)*}_b I_b \left( \bar{n}_B \hat{L}_\alpha \otimes [b] + n_B [b] \otimes L^\dagger_{\alpha} \right)
  \right. \nonumber \\ \left.
  + \Phi^{(2)}_\alpha I_\alpha \left( \bar{n}_B [\alpha]\otimes \hat{L}_b + n_B \hat{L}^\dagger_b \otimes [\alpha] \right) \right. \nonumber \\
  \left. \bar{n}_B \tilde{\kappa}_{\alpha a} [\beta] \hat{L}_a \otimes \hat{L}_b 
  -n_B \tilde{\kappa}_{\beta b} [b] \hat{L}^\dagger_\beta \otimes \hat{L}^\dagger_\alpha 
  \right. \nonumber \\ \left.
  -\bar{n}_B \tilde{\kappa}^*_{\beta b }  \hat{L}_\alpha \otimes \hat{L}_\beta [b]
  +n_B \tilde{\kappa}^*_{\alpha a}  \hat{L}^\dagger_b \otimes \hat{L}^\dagger_a [\beta]\right) \nonumber
\end{align}
Here we use short-hand notations $[a] = [\hat{L}^\dagger_a, \hat{L}_a]$, $\bar{n}_B \equiv 1+n_B$.
The terms not containing $\Phi$ can be written alternatively as 
\begin{align}
(A_+ A_- - A_- A_+)_3 = i \times \\
\bar{n}_B \left(L_b \otimes L_a\left( 
\gamma_{a\alpha}\tilde{\kappa}_{\alpha b} ([\alpha] \otimes 1 ) + \tilde{\kappa}^*_{a \alpha} \gamma_{\alpha b} (1 \otimes [\alpha])
\right) \right. \nonumber \\ \left.
-  L_\beta \otimes L_\alpha \left(
\gamma_{\alpha a} \tilde{\kappa}_{a \beta}  ([a] \otimes 1 ) + \tilde{\kappa}^*_{\alpha a} \gamma_{a \beta} (  1\otimes [a] )
\right) \right) \nonumber \\
+n_B \left(L^\dagger_{a} \otimes L^\dagger{b} \left( 
\tilde{\kappa}_{a\alpha} \gamma_{a\beta} ([\alpha] \otimes 1 ) + \gamma_{a \alpha}\tilde{\kappa}^*_{\alpha b}
\right) \right. \nonumber \\ \left.
- L^\dagger_{\alpha} \otimes L^\dagger_{\beta}\left( 
\tilde{\kappa}_{\alpha a} \gamma_{a\beta} ([a] \otimes 1 ) + \gamma_{\alpha a }\tilde{\kappa}^*_{a \beta} (1 \otimes [a])
\right) \right) \nonumber
\end{align}

If we disregard dissipative parts of environment susceptibilities, $\gamma_{a\beta} =0$,
we can restrict ourselves to $(A_+ A_- - A_- A_+)_2$. Generally, we have to work with all terms.

Let us simplify assuming that all "Latin" systems are oscillators and all "Greek" systems are qubits, so that $[\hat{L}^\dagger_a, \hat{L}_a] = -1$, 
$[\hat{L}^\dagger_\alpha, \hat{L}_\alpha] = \tau^z_\alpha$. We replace $ \hat{L}_a \to \hat{b}_a$; $\hat{b}_a$ being the annihilation operators in the oscillators, $\hat{L}_\beta; \hat{L}^\dagger_\beta \to \sigma^{\mp}_\beta$.  All the terms can be collected as follows:
\begin{align}
\partial_t \hat{\rho} = -i [(\hat{H}_{\rm a} +\hat{H}_{\rm b}),\hat{\rho}] +
(\partial_t \hat{\rho})_{\rm c} + (\partial_t \hat{\rho})_{\rm d}+ (\partial_t \hat{\rho})_{\rm e}.
\end{align}
Two contributions are purely Hamiltonian and contain oscillating fluxes,
\begin{align}
\hat{H}_{\rm a} = -|\Phi^{(2)}_\alpha|^2 \frac{I_\alpha^2}{2} ;\label{eq:quadratic}\\
\hat{H}_{\rm b} = I^2_a I_\beta \left( -(\Phi^{(1)}_a \chi_{\beta a} \hat{\sigma}^+_\beta + \Phi^{(1)*}_a \chi^*_{\beta a} \hat{\sigma}^-_\beta \right). \label{eq:phi1}
\end{align}

The contribution $({\rm c})$ describes the effect of $\Phi^{(2)}$ on the qubits and is not purely Hamiltonian,
\begin{align}
(\partial_t \hat{\rho})_{\rm c} \Omega = i \frac{I_{\alpha}}{2} \left[ \left(\Phi^{(2)*}_\alpha \kappa_{\alpha a} \hat{b}_a + \Phi_\alpha \kappa_{a \alpha} \hat{b}_a^\dagger\right)\tau^z_\alpha, \hat{\rho}\right] \nonumber \\
+ I_\alpha \left( \bar{n}_B (\Phi^{(2)*}_\alpha \gamma_{\alpha a} {\cal D}(\hat{b}_a,\hat{\rho},\tau^z_\alpha) + \Phi_\alpha^{(2)}\gamma_{a \alpha} {\cal D}(\tau^z_\alpha,\hat{\rho},\hat{b}^\dagger_a)) 
\right. \nonumber \\ \left. 
+ n_B(\Phi^{(2)*}_\alpha \gamma_{\alpha a} {\cal D}(\tau^z_\alpha,\hat{\rho},\hat{b}_a) + \Phi_\alpha^{(2)}\gamma_{a \alpha} {\cal D}(\hat{b}^\dagger_a,\hat{\rho},\tau^z_\alpha) ) \right)
\end{align}

The contribution $({\rm d})$ comprises the qubit operators only
\begin{align}
(\partial_t \hat{\rho})_{\rm d} \Omega = \frac{i}{8} \left(\kappa_{\alpha a} \kappa_{a \beta} - \gamma_{\alpha a}\gamma_{a \beta} \right) \left[\{\hat{\sigma}^-_\alpha,\hat{\sigma}^+_\beta\},\hat{\rho}\right] + \nonumber \\
\frac{\kappa_{\alpha a} \gamma_{a \beta}+ \gamma_{\alpha a} \kappa_{a \beta}} {2} \left( \bar{n}_B{\cal D}(\hat{\sigma}^-_\alpha,\hat{\rho},\hat{\sigma}^+_\beta) +n_B({\cal D}(\hat{\sigma}^+_\beta,\hat{\rho},\hat{\sigma}^-_\alpha))\right  )
\label{eq:qubit-only}
\end{align}

Finally, the most interesting contribution$({\rm e})$ describes the coupling of the qubits and the oscillators
\begin{align}
(\partial_t \hat{\rho})_{{\rm e}} \Omega = \frac{i}{8} \left(\kappa_{ a\alpha} \kappa_{\alpha b} - \gamma_{ a\alpha}\gamma_{\alpha b}\bar{n}_B^2\right) \left[\tau^z_\alpha \{\hat{b}^\dagger_a,\hat{b}_b\},\hat{\rho}\right] \nonumber \\ 
+\frac{1}{4} \left(\kappa_{ a\alpha} \gamma_{\alpha b} + \gamma_{ a \alpha}\kappa_{\alpha b}\right)\left(\bar{n}_B ({\cal D}(\hat{b}_b \tau^z_\alpha, \hat{\rho}, \hat{b}^\dagger_a)+ {\cal D}(\hat{b}_b , \hat{\rho}, \tau^z_\alpha\hat{b}^\dagger_a)) \right. \nonumber \\ \left.
+ n_B ({\cal D}(\hat{b}^\dagger_a \tau^z_\alpha, \hat{\rho},\hat{b}_b )+ {\cal D}( \hat{b}^\dagger_a, \hat{\rho}, \tau^z_\alpha\hat{b}_b)) \right)  \nonumber \\ 
+\frac{i}{2} \gamma_{ a\alpha}\gamma_{\alpha b} \left( (\bar{n}_B+n_B) \left(\bar{n}_B {\cal D}(\hat{b}_b, [\tau^z_\alpha, \hat{\rho}], \hat{b}^\dagger_a) \right. \right.\nonumber \\ \left. \left.
+ n_B {\cal D}(\hat{b}^\dagger_a, [\tau^z_\alpha, \hat{\rho}],\hat{b}_b)\right)  + \bar{n}_B n_B \left(  \{\hat{b}^\dagger_a,\hat{b}_b\}\hat{\rho}\tau^z_\alpha - \tau^z_\alpha \hat{\rho}\{\hat{b}^\dagger_a,\hat{b}_b\} \right)
\right) \label{eq:acstark-3}
\end{align}

To make sense out of this, let us compute the change of the qubit response functions owing to the presence of oscillators, and, correspondingly, the change of the oscillator response functions in the presence of the qubits. Note that the inverse susceptibilities (those related to admittances) of the oscillators/qubits in the frequency range of interest read:
\begin{equation}
A_{a} = - \frac{I^2_a}{\omega - f_2 + i\delta}; \; A_{\alpha} = \frac{I^2_\alpha \tau_z^\alpha}{\omega - f_1 + i\delta};
\end{equation}
the qubit admittances being state-dependent. The factor $\Omega^{-1}$ eventually stems from the admittances. 
With this, the qubit-qubit response function (at $\omega \approx f_1$ )is modified as
\begin{align}
\chi_{\alpha \beta} \to \chi_{\alpha \beta} + \delta \chi_{\alpha \beta}, \nonumber \\
 \delta \chi_{\alpha \beta} \equiv \chi_{\alpha a} A_{a} \chi_{a\beta} = \chi_{a \alpha} \frac{I^{2}_\alpha}{\Omega}\chi_{\alpha b}
\end{align}
while the oscillator-oscillator response function (at $\omega \approx f_2$ ) is modified as
\begin{align}
\chi_{ab} \to \chi_{ab} + \delta \chi_{a b}, \nonumber \\
\delta \chi_{a b} \equiv \chi_{a \alpha} A_{a} \chi_{\alpha b} = \chi_{a \alpha} \frac{I^{2}_\alpha \tau^z_\alpha}{\Omega}\chi_{\alpha b}
\end{align}
and depends on the state of the qubits.

Let us start with the interpretations. We rewrite the Hamiltonian term $\hat{H}_{\rm a}$ given by Eq.~\eqref{eq:quadratic} as
\begin{align}
H_{\rm a} = - |\Phi^{(2)}_\alpha|^2 \frac{I_\alpha^2}{\Omega} \tau_z = - \Phi^{(2)*}_\alpha A_\alpha \Phi^{(2)}_\alpha = - \Phi^{(2)*}_\alpha I_\alpha^{(2)}, \label{eq:nonlinearity}
\end{align}
where $I_\alpha^{(2)} \equiv A_\alpha \Phi_\alpha^{(2)}$ is the average current in qubit $\alpha$ induced by the non-resonant drive on this qubit. The term thus describes the effect of non-resonant drive on the qubit polarization that is quadratic in the drive amplitude, that is, manifests non-linearity of the qubits. As we will see, this term will help us to understand several others.

The Hamiltonian term $\hat{H}_{\rm b}$ (Eq.~\eqref{eq:phi1})
\begin{align}
\hat{H}_{\rm b}= - \frac{I^2_a I_\beta}{\Omega} (\Phi^{(1)}_a \chi_{\beta a} \hat{\sigma}^+_\beta + \Phi^{(1)*}_a \chi^*_{\beta a} \hat{\sigma}^-_\beta)
\end{align}
has the same form as the qubit coupling with $\Phi^{(1)}_\beta$. Thus it can be regarded as the addition to the drives at the qubits $\Phi^{(1)}_\beta$ owing to the currents in the oscillators,
\begin{align}
\Phi^{(1)}_\beta \to \Phi^{(1)}_\beta + \chi_{\beta a} A_a \Phi^{(1)}_a
\end{align}

In distinction from this, the term $(\partial_t \hat{\rho})_{\rm c}$ proportional to $\Phi^{(2)}_a$  is not purely Hamiltonian causing decoherence for non-diagonal components of density matrix. To understand, let us establish two limits where the term becomes purely Hamiltonian. Let us first assume that $\hat{\rho}$ commutes with all $\tau^{z}_\alpha$, that is, the state of the qubits is fixed. The term $(\partial_t \hat{\rho})_{\rm c}$ is then reduced to a Hamiltonian 
\begin{align}
\hat{H}^{\rm Q}_{\rm c} = -\frac{I^2_{\alpha}\tau^z_\alpha}{\Omega} I_a (\Phi^{(2)}_\alpha \chi_{a \alpha }\hat{b}_a^\dagger + \Phi^{(2)*}_\alpha \chi^*_{a \alpha }\hat{b}_a ).
\end{align}
It has the same form as the coupling of the oscillator current with the corresponding flux $\Phi^{(2)}_a$ and thus represents the addition to the fluxes in the oscillators $\Phi^{(2)}_a$ owing to the currents in the qubits,
\begin{align}
\Phi^{(2)}_a \to \Phi^{(2)}_a  + \chi_{ a \alpha} A_\alpha \Phi^{(2)}_\alpha
\end{align}

To establish the second limit, let us assume that  $\hat{\rho}$ commutes with all $\hat{b}_\alpha$, that is, the state of radiation. The term is reduced to a {\it different} Hamiltonian
\begin{align}
\hat{H}^{\rm O}_{\rm c} = -\frac{I^2_{\alpha} \tau^z_\alpha}{\Omega} I(\Phi^{(2)}_\alpha \chi^*_{\alpha a  }\hat{b}_a^\dagger + \Phi^{(2)*}_\alpha \chi_{\alpha a }\hat{b}_a )
\label{eq:phi2-8}
\end{align}
It can be interpreted if we invoke the qubit non-linearity (Eq.~\eqref{eq:quadratic}) and notice that the currents in the oscillators induce flux on the qubits. Thus this Hamiltonian term describes addition to  $\Phi^{(2)}_\alpha$ in the non-linear term due to the currents $\tilde{I}_a$ in the oscillators,
\begin{align}
\label{eq:addition}
\Phi^{(2)}_\alpha  \to \Phi^{(2)}_\alpha + \chi_{\alpha a} \tilde{I}_a ; \tilde{I}_a \equiv I_a \hat{b}_a. 
\end{align}

The contribution $({\rm d})$ given by Eq.~\eqref{eq:qubit-only} depends on the qubit degrees of freedom only and as such is expressed via the change of the qubit-qubit response function $\delta \chi_{\alpha \beta}$ owing to the presence the oscillators. Introducing
\begin{align}
\delta {\lambda}_{\alpha \beta} = - \frac{1}{2} (\delta \chi_{\alpha\beta} + \delta \chi^*_{\beta\alpha}) I_\alpha I_\beta, \label{eq:deltalambda} \\ \delta \Gamma_{\alpha \beta} = -i I_\alpha I_\beta \left(\delta \chi_{\alpha\beta}- \delta \chi^*_{\beta\alpha}\right) \label{eq:deltagamma}
\end{align}
we arrive at (c.f. Eq.~\eqref{eq:secondorder})
\begin{align}
(\partial_t \hat{\rho})_{\rm d}   = \frac{i}{2} \delta {\lambda}_{\alpha \beta} \left[\{\hat{\sigma}^-_\alpha,\hat{\sigma}^+_\beta\},\hat{\rho}\right] +  \nonumber \\
\delta G_{\alpha\beta} \left( \bar{n}_B{\cal D}(\hat{\sigma}^-_\alpha,\hat{\rho},\hat{\sigma}^+_\beta) 
  +n_B({\cal D}(\hat{\sigma}^+_\beta,\hat{\rho},\hat{\sigma}^-_\alpha))\right  )
\end{align}

Finally, let us discuss the most important contribution $(\rm e)$ (Eq.~\eqref{eq:acstark-3}): it describes the ac Stark effect, interaction between the qubits and radiation in the oscillators. Its general form is rather involved. To comprehend it better, we  consider again two limits. If $\hat{\rho}$ commutes with $\tau^z_\alpha$, so the state of the qubits is fixed, the contribution $(\rm e)$  is expressed via the change of the oscillator-oscillator response function, this change does depend on the state of the qubits. In this limit, the contribution thus become
\begin{align}
(\partial_t \hat{\rho})^{\rm Q}_{\rm d} = -i \left[ \hat{H}^{\rm Q}_{\rm e},  \hat{\rho}\right] + \nonumber \\
\delta \Gamma_{ab}
\left(\bar{n}_B {\cal D}(\hat{b}_b , \hat{\rho}, \hat{b}^\dagger_a)
+ n_B ({\cal D}(\hat{b}^\dagger_a , \hat{\rho},\hat{b}_b )\right)  \\
\hat{H}^{\rm Q}_{\rm e} = - \frac{1}{2} \delta \lambda_{ab} \{\hat{b}^\dagger_a,\hat{b}_b\} \nonumber \\ = - \frac{I_a I^2_\alpha I_b}{\Omega}   \frac{\chi_{a \alpha} \chi_{\alpha b} + \chi^*_{b \alpha} \chi^*_{\alpha a}} {2} \tau_\alpha^{z} \{\hat{b}^\dagger_a,\hat{b}_b\} 
\label{eq:fixedq}
\end{align}
where we define the state-dependent quantities $\delta {\lambda}_{ab}$, $\Gamma_{ab}$ similar to Eqs. (\ref{eq:deltalambda}),(\ref{eq:deltagamma}).

In the opposite limit $\hat{\rho}$ commutes with $\hat{b},\hat{b}^\dagger$, so that the state of radiation is fixed. The contribution $(\rm e)$ is reduced to a Hamiltonian
\begin{align}
\hat{H}_{\rm e}^{\rm O}= - \frac{I_a I_\alpha^2 I_b}{\Omega} \tau^z_\alpha \{\hat{b}^\dagger_a,\hat{b}_b\} \chi_{\alpha b} \chi^*_{\alpha a}
\end{align} 
The way to understand this Hamiltonian is the same as we use for $\hat{H}_{\rm c}^{\rm O}$ (Eq.~\eqref{eq:phi2-8}):  it stems from the non-linearity term ( Eq.~\eqref{eq:quadratic}) with the addition to the fluxes given by Eq.~\eqref{eq:addition}. 

In conclusion, let us mention that all the relevant terms become Hamiltonian if we disregard the dissipative part of the susceptibility, so that 
$\chi_{\alpha \beta} = \chi^*{\beta \alpha}$. The Hamiltonian limits discussed become the same, $\hat{H}^{\rm Q}_{{\rm d},{\rm e}} = \hat{H}^{\rm O}_{{\rm d},{\rm e}}$. The contribution $({\rm d})$ is expressed thorough the modification of the susceptibility, and contributions $({\rm a})$, $({\rm c})$, $({\rm e})$ are collected to the single expression that describes the non-linear response of each qubit on the flux including a part contributed by the oscillators:
\begin{align}
\hat{H}_{\rm a} + \hat{H}_{\rm c} +\hat{H}_{\rm e} = - \frac{I_\alpha^2}{2 \Omega} \{\hat{\Phi}^{(2) \dagger}_\alpha, \hat{\Phi}^{(2)}_\alpha \} \; \\
\hat{\Phi}^{(2)}_\alpha  = \Phi^{(2)}_\alpha + \chi_{\alpha a} I_a \hat{b}_a
\end{align}

\section{\label{sec:steady_state_interaction}Solution of the master equation for weakly interacting driven qubits}

In this appendix we explicitly solve the steady state Lindblad equation Eq.~\eqref{eq:Lindblad} in the small $\lambda$ limit at the crossing point where the frequencies of the dressed ALQs are equal $\xi_1=\xi_2$. We use this solution to obtain Eq.~\eqref{eq:peak_z1} in the main text.  The exact assumption we make is that $\lambda$ is small compared to $\xi_{1, 2}$ at the crossing point. In particular this holds if both the drive frequency $f$ and either of the drive amplitude $A_{1,2}$ is large compared to $\lambda$.

For $\lambda{=}0$ the effective Hamiltonian is diagonalized by the dressed ALQ states
\begin{align}\label{eq:dressed_states}
    \ket{e'e'} &= U_1 U_2 \ket{ee} \quad \quad \ket{e'g'} = U_1 U_2 \ket{e g} \nonumber \\
    \ket{g'e'} &= U_1 U_2 \ket{ge} \quad \quad \ket{g'g'} = U_1 U_2 \ket{gg}, 
\end{align}
with corresponding energies 
\begin{align}
    E_{e'e'} &= \xi_1 + \xi_2 \quad \quad E_{e' g'} = \xi_1 - \xi_2 \nonumber \\
    E_{g' e'} &= \xi_2 - \xi_1  \quad \quad E_{g'g'} = -\xi_1 - \xi_2, 
\end{align}
where
\begin{equation}\label{eq:U}
    U_{\alpha}= e^{\frac{\eta_{\alpha}}{2} \sigma_{\alpha}^x \sigma_{\alpha}^z}, \quad \cos(\eta_{1, 2}) = \frac{f \pm \varphi}{\xi_{1, 2}},
\end{equation}
and $\xi_{1, 2}=\sqrt{(f{\pm} \varphi)^2{+}A_{1, 2}^2}$.
The interaction between the dressed qubits becomes significant for $|\xi_1 - \xi_2| \lesssim |\tilde{\lambda}|$, where 
\begin{align}\label{eq:lambda_tilde}
    \tilde{\lambda} &\equiv \bra{g'e'} H_{\text{eff}}\ket{e'g'}\nonumber\\
    &= \frac{i \lambda}{4} \Bigl(e^{i \theta}  (1 - \cos(\eta_1))(1- \cos(\eta_2)) \nonumber \\
    &- e^{-i\theta} (1 + \cos(\eta_1))(1+ \cos(\eta_2))\Bigr)
\end{align}
is the overlap between dressed states.
To lowest order in $\lambda/(\xi_{1} + \xi_2)$ we account for the interaction by performing an additional rotation between $\ket{e'g'} $ and $\ket{g'e'}$ to obtain the the eigenstates $\ket{1} = \ket{e'e'}$, $\ket{4}=\ket{g'g'}$,
\begin{align}
    \ket{2} &= \cos\left(\frac{\eta_3}{2}\right)\ket{e'g'} + \frac{\tilde{\lambda}}{|\tilde{\lambda}|}\sin\left(\frac{\eta_3}{2}\right)\ket{g'e'},\\
    \ket{3} &=  \cos\left(\frac{\eta_3}{2}\right)\ket{g'e'} - \frac{\tilde{\lambda}^*}{|\tilde{\lambda}|}\sin\left(\frac{\eta_3}{2}\right)\ket{e'g'},
\end{align} 
where
\begin{equation}
    \cos\left(\eta_3\right) = \frac{\xi_1 - \xi_2}{\sqrt{(\xi_1 - \xi_2)^2 + |\tilde{\lambda}|^2}}.
\end{equation} 
By inserting in Eq.~\eqref{eq:rates} we obtain the non-zero rates
\begin{align}
    \Gamma_{3\leftarrow 1} &= \Gamma_{4 \leftarrow 2} = \Gamma^{+}_{(12)} + \cos(\eta_3) \Gamma^{+}_{[12]}\\
    \Gamma_{2\leftarrow 1} &= \Gamma_{4 \leftarrow 3} =\Gamma^{+}_{(12)} - \cos(\eta_3) \Gamma^{+}_{[12]} \\
    \Gamma_{1\leftarrow 3} &= \Gamma_{2 \leftarrow 4} =\Gamma^{-}_{(12)} + \cos(\eta_3) \Gamma^{-}_{[12]}\\
    \Gamma_{1\leftarrow 2} &= \Gamma_{3 \leftarrow 4} =\Gamma^{-}_{(12)} - \cos(\eta_3) \Gamma^{-}_{[12]}\\
    \Gamma_{2\leftarrow 3} &= \Gamma_{3 \leftarrow 2} = \sin^2(\eta_3) \Gamma_\lambda,
\end{align}
where 
\begin{equation}
    \Gamma_{\lambda} =  \frac{1}{2}(\gamma_1 \sin^2(\eta_1) + \gamma_2 \sin^2(\eta_2)),
\end{equation}
\begin{equation}
    \Gamma^{\pm}_{(12)} = \frac{\Gamma^{\pm}_{1} + \Gamma^{\pm}_{2}}{2}, \quad \Gamma^{\pm}_{[12]} = \frac{\Gamma^{\pm}_{1} - \Gamma^{\pm}_{2}}{2},
\end{equation}
and 
\begin{equation}
    \Gamma_\alpha^\pm = \frac{\gamma_\alpha}{2}\left(1 \pm \cos(\eta_\alpha) \right)^2.
\end{equation}
Inserting in the master equation Eq.~\eqref{eq:master} we obtain the steady state populations
\begin{align}
    N\rho_1 &= 2s_3^2 \Gamma_\lambda \Gamma_{(12)}^- \Gamma_{(12)}^- + ( \Gamma^{-}_{(12)} + c_3 \Gamma^{-}_{[12]})\nonumber\\\
    &\times( \Gamma^{-}_{(12)} - c_3 \Gamma^{-}_{[12]}) (\Gamma^{-}_{(12)}  + \Gamma^{+}_{(12)} )\\
    N\rho_2 &=2s_3^2 \Gamma_\lambda \Gamma_{(12)}^- \Gamma_{(12)}^+ + ( \Gamma^{-}_{(12)} + c_3 \Gamma^{-}_{[12]})\nonumber\\
    &\times( \Gamma^{+}_{(12)} - c_3 \Gamma^{+}_{[12]})(\Gamma^{-}_{(12)}  + \Gamma^{+}_{(12)} )\\
    N\rho_3 &= 2s_3^2 \Gamma_\lambda \Gamma_{(12)}^- \Gamma_{(12)}^+ + ( \Gamma^{+}_{(12)} + c_3 \Gamma^{+}_{[12]})\nonumber\\
    &\times( \Gamma^{-}_{(12)} - c_3 \Gamma^{-}_{[12]}) (\Gamma^{-}_{(12)}  + \Gamma^{+}_{(12)} )\\
    N\rho_4 &=2 s_3^2 \Gamma_\lambda \Gamma_{(12)}^+ \Gamma_{(12)}^++ ( \Gamma^{+}_{(12)} + c_3 \Gamma^{+}_{[12]})\nonumber\\
    &\times( \Gamma^{+}_{(12)} - c_3 \Gamma^{+}_{[12]}) (\Gamma^{-}_{(12)}  + \Gamma^{+}_{(12)} )
\end{align}
where we introduced $c_i=\cos(\eta_i)$, $s_i=\sin(\eta_i)$, and $N$ is defined by $\sum_i \rho_i=1$.

Exactly at the crossing $c_3=0$, and by using $\langle \sigma_{1}^z \rangle = c_{1}(\rho_1{-}\rho_4) \pm c_{1} c_3 (\rho_3-\rho_4)$ we obtain 
\begin{align}
    \langle \sigma_1^z \rangle &= c_1 \frac{\Gamma^-_{(12)} - \Gamma^+_{(12)}}{\Gamma^-_{(12)} + \Gamma^+_{(12)}}\nonumber \\
     &= -2c_1\frac{c_1\gamma_{1} +c_2\gamma_2}{\gamma_1(1 + c_1^2) + \gamma_2(1 + c_2^2)}
\end{align}
which yields Eq.~\eqref{eq:peak_z1} after inserting Eq.~\eqref{eq:U} with $\varphi=\varphi_c$.

\end{document}